\documentclass[aps,prl,twocolumn,superscriptaddress,floatfix]{revtex4-2}
\usepackage{graphicx}
\usepackage{amsmath}
\usepackage{amssymb}
\usepackage{xcolor}
\usepackage{siunitx}
\usepackage[newitem,newenum]{paralist}
\usepackage{hyperref}

\usepackage{booktabs}
\usepackage[font=small,justification=raggedright,singlelinecheck=false]{caption}
\usepackage[margin=false,inline=true,index=true,draft]{fixme}
\fxusetheme{colorsig}

\hypersetup{hidelinks}
\newcommand{\myparagraph}[1]{\par\textbf{#1}}
\newcommand{\dm}{{\text{\sc dm}}}
\newcommand{\added}[1]{#1}

\newcommand{\ion}[2]{#1\,\textsc{\lowercase{#2}}}

\usepackage[normalem]{ulem}

\usepackage{booktabs}
\usepackage{xspace}
\usepackage{xr}

\def\resolve{Resolve\xspace}

\newcounter{smsection}
\newcounter{smsubsection}[smsection]

\renewcommand{\thesmsection}{S\arabic{smsection}}
\renewcommand{\thesmsubsection}{\thesmsection.\arabic{smsubsection}}

\newcommand{\smsection}[2]{%
  \refstepcounter{smsection}%
  \section*{\thesmsection\quad #1}%
  \label{#2}%
}

\newcommand{\smsubsection}[2]{%
  \refstepcounter{smsubsection}%
  \subsection*{\thesmsubsection\quad #1}%
  \label{#2}%
}

\begin{document}

\title{Recycling serendipitous observations of XRISM for dark matter searches}
\author{Lidiia Zadorozhna}
\affiliation{Niels Bohr Institute, University of Copenhagen, Denmark}
\affiliation{Taras Shevchenko National University of Kyiv, Ukraine}
\author{Denys Malyshev}
\affiliation{Institut f{\"u}r Astronomie und Astrophysik T{\"u}bingen, Universit{\"a}t T{\"u}bingen, Germany}
\author{Oleg Ruchayskiy}
\affiliation{Niels Bohr Institute, University of Copenhagen, Denmark}
\begin{abstract}
Searches for radiative decays of keV-mass dark matter have been limited by energy resolution.
We propose a strategy for the XRISM/Resolve low count rate observations, and search the $3.48$--$3.56$~keV band for a decay line from the Milky Way halo. 
Despite a lower exposure, our bounds are stronger than those from XRISM observations of galaxy clusters.
We also identify a residual at $E = 3.5311$~keV with $\sim 3\sigma$ local significance, coincident with the previously reported unidentified feature, and having no plausible atomic or instrumental counterpart. 
These bounds will continue to tighten as additional pointings become available.
\end{abstract}

\maketitle

The identification of the nature of dark matter (DM) remains one of the most important open problems in fundamental physics. 
Although DM is commonly pictured as a stable particle, absolute stability is not required by any observation: a decay lifetime far exceeding the age of the Universe is fully consistent with all cosmological and astrophysical constraints, making metastable DM a viable and well-motivated hypothesis.
If the dominant decay channel is radiative, the resulting two-body kinematics imprints a narrow, monochromatic line on the diffuse X-ray (or $\gamma$-ray) sky at energy $E_\gamma= m_{\dm}/2$. 

X-ray searches for decaying DM~\cite[see, e.g.,][]{Abazajian:2001vt,Boyarsky:2006fg,Boyarsky:2018tvu} have been limited by spectral resolution.
The line's broadening is set entirely by the virial velocities of the host halo and ranges from $\Delta E/E \sim\,10^{-4}$ in dwarf spheroidal galaxies to $10^{-2}$ in galaxy clusters~\cite{2009arXiv0906.1788D}.
The CCD-class instruments (XMM-Newton, Chandra, Suzaku) that have carried the program until recently delivered only $\Delta E/E \lesssim  10^{-1}$~\cite{2001A&A...365L...1J,2007PASJ...59S...1M,2000SPIE.4012....2W}. 
As a result, a clean separation of a candidate line from the continuum and from neighboring astrophysical lines was ambiguous, making searches for weak unidentified features challenging.

This ambiguity manifested itself in the case of the weak unidentified spectral feature at energies $E \approx 3.5$~keV reported initially in~\cite{Bulbul:2014sua,Boyarsky:2014jta} (and subsequently confirmed by many groups~\cite{2015PhRvL.115p1301B,2015arXiv150805186I,2016ApJ...829..124F,2015MNRAS.451.2447U,Neronov:2016wdd,Cappelluti:2017ywp,2018arXiv181210488B,2019A&A...625L...7H}; see, however,~\cite{2014PhRvD..89b5017H, 2014PhRvD..90j3506M, Riemer-Sorensen:2014yda,2015PASJ...67...23T, XQC:2015mwy, 2016PASJ...68S..31S, 2016A&A...592A.112H,Sicilian:2020glg}).
A decade of analysis has left two controversies unresolved. 
The first one is the \emph{detection ambiguity}.
At keV energies, atomic lines are closely spaced, with typical separations smaller than the spectral resolution of a CCD detector. 
A single resolution element may therefore contain several lines, which blend into a single broadened feature. Since the individual lines cannot be resolved, the flux attributed to any candidate line is determined jointly with the local continuum, and therefore depends sensitively on the assumed continuum model.
For example, a stack of XMM-Newton blank sky data was claimed to exclude the presence of the line~\mbox{\cite{Dessert:2018qih,Dessert:2023fen}}. This result was, however, contested in Refs.~\mbox{\cite{Boyarsky:2020hqb,Abazajian:2020unr}}, which demonstrated that narrow band modeling may lead to confusion between a real line and a continuum, thus hiding the spectral feature.

The second controversy concerns the \emph{nature} of the line. 
Even if some excess is present, it might be of astrophysical origin, e.g., \ion{K}{XVIII} or \ion{Cl}{XVII} lines~\cite{Foster:2012hy,Jeltema:2014qfa,Carlson:2014lla}; see, however, Ref.~\cite{Boyarsky:2014paa}. 
Another explanation via a charge-exchange emission from sulfur ions was proposed in~\cite{Gu:2015gqm}.
Both controversies ultimately arise from insufficient spectral resolution; the relevant systematic effects are reviewed in detail in~\cite{Boyarsky:2018tvu,Abazajian:2017tcc}.

The observation of the Perseus galaxy cluster by the Hitomi mission~\cite{Hitomi:2016mun} showed no excess at the energies of interest, thus disfavoring its astrophysical origin.
Indeed, thermal ion velocities in the intracluster medium ($\sim\,150~\mathrm{km\,s^{-1}}$) are far below the dark matter virial dispersion ($\sim\,1000~\mathrm{km\,s^{-1}}$). An astrophysical line of the reported intensity would therefore be narrow and should have been resolved by Hitomi, as were other subsolar abundance lines in~\cite{Hitomi:2016mun}.
The dark matter interpretation, by contrast, remained unconstrained: the exposure was insufficient to detect a weak line spread over several bins~\cite{Boyarsky:2018tvu,Abazajian:2017tcc,Iakubovskyi:2014yxa}.

The microcalorimeter \resolve\ on board the X-Ray Imaging and Spectroscopy Mission (XRISM)~\citep{Tashiro:2022ful,2020arXiv200304962X}, launched
in September 2023, provides resolving power identical to that of
Hitomi ($\mathrm{FWHM} \simeq 5$~eV at 3.5~keV
~\citep{xrism_pog}) with a
viable mission lifetime, motivating dedicated sensitivity forecasts
for decaying DM searches~\citep{Dessert:2023vyl,Zhou:2024sto}. 
A first dedicated search for a DM decay line in observations of galaxy clusters~\citep{XRISM:2025lzv} could not fully exploit the resolution advantage. For a massive cluster with the virial velocity 
$v_{\rm vir} \sim\,1000$~km\,s$^{-1}$, the DM line is Doppler broadened over several resolution elements and must be identified on top of a strong astrophysical continuum. 

\emph{In this work} we propose a line search strategy optimized for \resolve. We recycle serendipitous observations with low count rates to search for the decay signal of DM in the Milky Way halo, turning the mission into a continuously accumulating DM survey~\cite{Boyarsky:2006fg,2009arXiv0906.1788D}.
This method has two advantages over the cluster searches.
First, the DM line is broadened with $v_{\rm vir} \simeq 220$~km\,s$^{-1}$, leading to a full width at half maximum (FWHM) a factor of ${\sim\,}5$ smaller than in galaxy clusters.
Second, in such observations, individual spectral channels contain zero to a few photons, and a potential line signal is \emph{comparable} to the local background per channel.
We refer to this as the \textit{low background regime}, not accessible to previous, lower resolution, higher background instruments (with the possible exception of Ref.~\citep{2002ApJ...576..188M}).

To demonstrate the capabilities of our approach, we show that already existing, publicly available low count rate observations (typically used for background estimation) are sufficient to produce the best constraints on decaying dark matter around 3.5~keV.

\myparagraph{Data selection.}
 The full per observation details
are provided in the Supplemental Material (SM)~\cite{SM}.

From all public \resolve fields available as of April 2026, we 
select the observations with the lowest astrophysical background as 
follows. We rank all observations with exposure time $T > 1000$~s by 
their total count rate $R$ of the archival cleaned
event list, exclude those that are within $10^\circ$ of the Galactic
Center, and select the $N_{\rm obs} = 60$ observations with the
lowest count rates (excluding two pointings with anomalous detector
pixels), up to $R = 0.1352$~cts\,s$^{-1}$, with total
exposure $T_{\rm tot} = 3.798$~Ms (ONTIME; the screened exposure
carried by the spectra is $T_{\rm exp} = 2.917$~Ms).

We extract the spectra from the whole field of view (FoV) following the standard XRISM data reduction
procedure~\citep{XRISM_ABC_Guide,Doyle2022_pipeline}.
At these count rates grade branching ratios and spectral resolution are unaffected~\citep{xrism_pog}. 
Fluxes are extracted from an effective circular aperture of radius 
$1.8'$ ($\Delta\Omega = 8.613\times10^{-7}$~sr), matched to the
aperture used to generate the ancillary response file (ARF; see SM, Sec.~S1).
The selected pointings have a low astrophysical background: 
the count rates in our sample, $R = 0.05$--$0.135$~cts\,s$^{-1}$, 
lie an order of magnitude below the bright source threshold, and the 
lowest of them approach the \resolve non-X-ray background 
level~\citep{xrism_pog}.

\myparagraph{Analysis.}
The extracted spectra are fitted jointly in \texttt{XSPEC}~\citep{Xspec} with a common 
\texttt{powerlaw}~$+$~\texttt{gaussian} model in the 
$3.43$--$3.57$~keV band: the power law represents the featureless 
astrophysical and instrumental background, and the Gaussian models the 
decaying DM signal. Each observation has an independent power law 
normalization, with the photon index frozen at $\Gamma_{\rm ph} = 
1.46$~\citep{1997MNRAS.285..449C,2002ApJ...576..188M}; varying $\Gamma_{\rm ph}$ over
the range of plausible background slopes typically changes the 
derived limits by less than $1\%$. The Gaussian has the width 
$\sigma_E = 2$~eV, consistent with the virial broadening in the 
Galactic halo.

Unlike in previous searches, at the \resolve energy resolution the 
motion of the observer can no longer be neglected. Therefore, the 
Gaussian in each observation is centered at the energy 
$E_{{\rm obs},i} = E_0 + \delta E_i$, where $\delta E_i$ is a
per observation Doppler correction to the expected DM line energy
$E_0$, accounting for the full velocity of the observer relative to 
the rest frame of the Galactic DM halo -- the solar velocity plus 
the Earth's heliocentric orbital velocity at the observation epoch, 
projected onto each line of sight (see SM, Sec.~S2 for details). Across our
sample the total Doppler shift spans $-2.7$ to $+2.9$~eV at 
3.5~keV, more than half the FWHM of the \resolve's line spread 
function ($\simeq 5$~eV). 

The relative normalizations of the Gaussians across observations are tied to 
the DM column density along the corresponding line of sight. For the center of the 
$i$-th pointing we compute the angle $\psi_i$ to the Galactic 
Center and evaluate the column density $S_{\dm}(\psi_i)$ for a 
Navarro--Frenk--White (NFW) halo profile with $r_s = \SI{20}{kpc}$ 
and the DM density at the Solar position $\rho_\odot = 
\SI{0.4}{GeV/cm^3}$. The Gaussian 
normalizations are then tied via the ratios 
$S_{\dm,i}/S_{\dm,\mathrm{ref}}$, where $S_{\dm,\mathrm{ref}} = 
1.948\times10^{22}~\mathrm{GeV\,cm^{-2}}$ is the column density of a 
reference pointing (ObsID 000114000).  
In what follows we call this method the \textit{joint fit}.

\begin{figure}[t]
\centering
\includegraphics[width=\columnwidth]{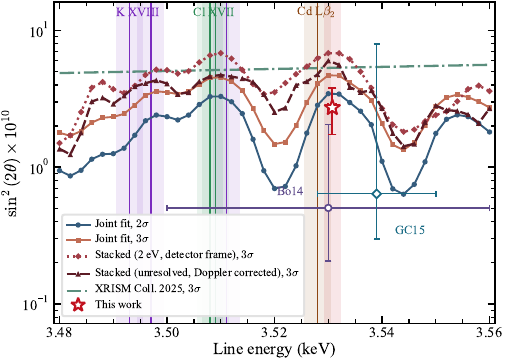}
\caption{Upper limits on the sterile neutrino mixing angle $\sin^2(2\theta)$ 
as a function of the line energy, derived from our \resolve\ sample 
($N_{\rm obs} = 60$, screened exposure $T_{\rm exp} = 2.917$~Ms). 
The blue and orange curves show the $2\sigma$ and $3\sigma$ limits 
from the \emph{joint fit}, respectively; the dotted red curve shows the $3\sigma$ limit from the \emph{stacked spectrum}, and the dark red dashed curve with triangles the same limit for an unresolved line in the Doppler corrected stack. The star marks 
the excess at $E \simeq 3.531$~keV with a local significance of
$3.0\sigma$ (Doppler corrected stack).
The red shaded region shows the $\pm 2\sigma$ uncertainty in the line position.
For comparison, the 
green dash-dotted curve shows the parametrized $3\sigma$ limit from 
the XRISM cluster stack~\citep{XRISM:2025lzv}, the purple point 
shows the unidentified signal reported in Ref.~\citep{Boyarsky:2014jta}
(M31), and the teal diamond the Galactic Center line of
Ref.~\citep{2015PhRvL.115p1301B}, its bar spanning the range of Milky
Way mass models considered there (the flux itself is measured to 17\%; the uncertainty of the Galactic dark matter profile dominates).
The vertical lines mark the nominal energies of known lines 
in this range: \ion{K}{xviii} (purple), 
\ion{Cl}{xvii} (green), and Cd~L$\beta_{2,15}$ (brown).
}
\label{fig:sin2_limits}
\end{figure}

As an alternative, we also consider a \textit{stacked fit} of all observations:
individual spectra are summed channel by channel, with ARF and redistribution matrix (RMF)
computed as an exposure weighted average.
We fit the resulting spectrum in \texttt{XSPEC} (C
statistic, unbinned) with the same \texttt{powerlaw}~$+$~\texttt{gaussian} model ($\sigma_E = \SI{2}{eV}$).
In this co-added spectrum a halo line sits at a different Doppler shifted energy
in every observation. We therefore also build a \textit{Doppler
corrected} stack, in which each spectrum is shifted to the rest frame
of the halo before the summation (SM, Sec.~S4); a halo line is then
aligned in all observations, whereas a line fixed in the detector
frame is smeared.

\myparagraph{Results.}
In this section we report the upper limits on the flux of a decaying 
DM line in the energy range $3.48$--$3.56$~keV. We parameterize the 
DM decay in terms of the mixing angle of the sterile neutrino, one 
of the most compelling decaying DM candidates~\citep{Dodelson:1993je}. 
The corresponding limits on the decay width and on the line flux are 
presented in the Supplemental Material.
For the stacked fit the flux limits are converted to DM parameters with the exposure weighted column density
$S_{\dm,\mathrm{eff}} = 2.13\times10^{22}~\mathrm{GeV\,cm^{-2}}$. 

The limits are shown in Figure~\ref{fig:sin2_limits}.  
The joint fit produces the limit that is stronger than the one obtained 
by the XRISM Collaboration from 3.75~Ms 
of observations of 10 clusters of galaxies~\cite{XRISM:2025lzv}.
The agreement between the joint and stacked limits validates the derived 
constraints.

Our stacked analysis reveals two line-like residuals above $2\sigma$   
(pre-trial) significance. 
The residual at $E = 3.5094 \pm 0.0018$~keV
(here and below we quote 68\% confidence level) coincides in energy with the blend of the \ion{Cl}{XVII} line 
at $3.5088$~keV and the \ion{K}{XVIII} line at 
$3.5105$~keV~\citep{Foster:2012hy}. 
In the stacked spectrum it
is detected at $2.2\sigma$ and has flux $F =
12.6^{+6.3}_{-5.9}\times10^{-8}~{\rm ph\,cm^{-2}\,s^{-1}}$. The 
interpretation of the 3.5~keV line as arising from these atomic 
transitions was discussed in \cite{Jeltema:2014qfa,Carlson:2014lla}.

\begin{figure}[t]
\centering
\includegraphics[width=\columnwidth]{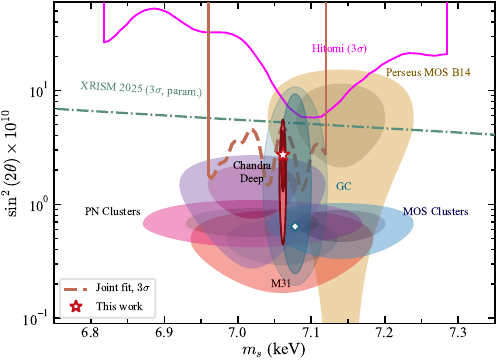}
\caption{\footnotesize The upper limit and the $3.531$~keV residual in the sterile neutrino parameter plane.
The dashed orange curve is the $3\sigma$
 limit of the joint fit, same as in Fig.~\ref{fig:sin2_limits}. 
The red star is the best fit of the residual in the Doppler corrected stack
with its $1\sigma$ and $2\sigma$  confidence
regions in dark and pale red. The shaded regions labeled in the plot
are the $1\sigma$ and $2\sigma$ regions (dark to pale) of
the earlier detections: the stacked XMM-Newton clusters and the core
removed Perseus spectrum~\citep{Bulbul:2014sua},
M31~\citep{Boyarsky:2014jta} and the Chandra deep
fields~\citep{Cappelluti:2017ywp}, Galactic Center line~\citep{2015PhRvL.115p1301B}.
The magenta and the teal dash-dotted
curves are the $3\sigma$ limits from the Hitomi observation of
Perseus~\citep{Hitomi:2016mun} and from the XRISM cluster
stack~\citep{XRISM:2025lzv}.
}
\label{fig:param_space}
\end{figure}

The second residual at energy $E = 3.5311^{+0.0010}_{-0.0014}$~keV  is detected with a
significance of $2.5\sigma$ ($F =
14.1^{+6.3}_{-5.9}\times10^{-8}~{\rm ph\,cm^{-2}\,s^{-1}}$). Its 
energy is fully consistent with that of the unidentified spectral feature 
reported in~\citep{Boyarsky:2014jta,Bulbul:2014sua,2015PhRvL.115p1301B}, see Fig.~\ref{fig:param_space}.
In the Doppler corrected stack (as detailed in SM, Sec.~S4) the residual reaches $3.0\sigma$ ($F = 13.5^{+5.3}_{-4.9}\times10^{-8}~{\rm ph\,cm^{-2}\,s^{-1}}$; $E = 3.5307^{+0.0007}_{-0.0008}$~keV).
The global significance is $2.0$--$2.5\sigma$ (SM, Sec.~S4).
This residual corresponds to ${26\pm 10}$ extra photons on top of $\approx 52$ background photons in a 5~eV bin.

The high spectral resolution 
of \resolve makes it possible to demonstrate that the residual 
is not associated with any known astrophysical or instrumental line. 
The low threshold non-equilibrium ionization (NEI) line list of AtomDB
v3.1.3~\citep{Foster:2012hy,2025AAS...24611804F}, used by the 
\texttt{apec} and \texttt{nei} models in \texttt{XSPEC}, contains no 
lines at this energy down to an emissivity threshold of 
$10^{-21}~{\rm ph\,cm^{3}\,s^{-1}}$. The published instrumental 
background~\citep{xrism_pog,XRISM:2025pca} contains no 
lines between $2.12$~keV and 
$5.4$~keV. Among the fluorescence lines of stable elements, the 
X-ray Data Booklet~\citep{KortrightThompson:XDB,1967RvMP...39...78B} lists two cadmium
transitions, Cd\,$L\beta_2$ and Cd\,$L\beta_{15}$, at $3.528$~keV.
The interpretation of the residual as Cd\,$L\beta_{2,15}$ is however excluded by the absence of the 
Cd\,$L\alpha_1$ line at $3.134$~keV, which would have to be 6.7 times 
stronger (SM, Sec.~S4).
We thus conclude that the residual at $E = 3.5311$~keV does not correspond to any known line.

Interpreted in terms of sterile neutrino DM, the residual corresponds to a best fit mixing angle of
\begin{equation}
\sin^2(2\theta) = (2.7 \pm 1.0)
\times10^{-10}
\label{eq:bestfit}
\end{equation}
(red
star in Figs.~\ref{fig:sin2_limits} and~\ref{fig:param_space}). 
It is compatible within
statistical uncertainties with previous detections~\citep{Bulbul:2014sua,Boyarsky:2014jta,2015PhRvL.115p1301B,Cappelluti:2017ywp}
and in $\approx
2\sigma$ tension with existing exclusions~\citep{2014PhRvD..90j3506M,2016PASJ...68S..31S,2016A&A...592A.112H}.

\myparagraph{Discussion.}
This work presents a strategy to search for decaying dark matter with XRISM/\resolve. The idea is that observations with low background, including those XRISM uses to estimate its background, can be used to search for a dark matter signal from the Milky Way halo. The strategy was advocated earlier~\cite{Boyarsky:2006fg} and has been used extensively with CCD X-ray telescopes, but the characteristics of XRISM require it to be reassessed.

Reusing 60 low background observations with a screened exposure of
$T_{\rm exp} = 2.92$~Ms, we derived bounds on a decaying DM line.
Our search is serendipitous. It requires no dedicated observations and uses data from the regular XRISM observing program.
Our limits are on average 70\% stronger than those from the XRISM observations of galaxy clusters~\citep{XRISM:2025lzv}, despite a lower
exposure and lower DM column density of our sample.
This demonstrates two advantages of our method.
First, the astrophysical background of the selected pointings is significantly lower than that of the 
cluster central fields. Second, the narrow Milky Way DM line 
fits within a single Resolve resolution element and is searched against a 
featureless continuum, whereas the cluster line is virially broadened
over several resolution elements and may be blended with thermal plasma emission 
lines nearby. 
The present bounds are also 
robust against systematic uncertainties. 
Our pointings lie
at angles $\psi_i > 38^\circ$ from the Galactic Center, where the DM
distribution of the Milky Way is well constrained (see Supplemental Material).

We found a positive residual at $E = 3.5311$~keV, the energy of 
the previously reported unidentified $3.5$~keV 
feature. 
The  residual does not correspond to any known astrophysical or 
instrumental line, although its global significance is only $2$--$2.5\sigma$, given the low photon statistics.

\textbf{Outlook.}
Low count rate observations are routinely accumulated in the XRISM 
observing program, typically as offset pointings needed to model
the background of bright targets, or as observations of low surface 
brightness fields~\citep{XRISM_LongTermPlan,XRISM_AO3_Targets}. The 
dataset thus grows as a byproduct of regular XRISM operations, at
no dedicated observing cost. The 60 observations analyzed here 
amount to $13\%$ of the $28$~Ms of 
XRISM exposure publicly available as of April 
2026~\citep{XRISM_DARTS_Public,XRISMMASTR}. We assume that low count 
rate fields keep supplying at least this fraction of the archival data. 
The mission is not limited to its nominal phase: the helium cryogenic 
cooling may last up to $6$~years~\citep{2025JATIS..11d2016P,Kelley:2025resolve}. 
At the current accumulation rate of $\sim 17$~Ms of exposure per year, 
the cryogen phase alone would double the archive to $\sim 100$~Ms by 
2029, bringing the low background sample to $\sim 13$~Ms, $3.4$ times 
the present one.

The factor of $3.4$ in exposure tightens the limits by a factor of
$\approx 1.8$. At $3.531$~keV this brings the $2\sigma$ upper limit to the upper edge of the range $\sin^2(2\theta)$ reported by~\citep{Boyarsky:2014jta}. 
The fate of the residual reported here will be
decided much sooner: in the same dataset a genuine line at the best
fit flux would reach $\approx 5\sigma$, whereas a background
fluctuation would fade. For comparison,  the cluster search needs a fivefold gain in sensitivity,
i.e.\ dedicated observing campaigns of $31.5$--$94$~Ms, to reach the
flux of the originally reported cluster
signal~\citep{XRISM:2025lzv,Bulbul:2014sua}.

The same low background will keep the analysis photon limited rather than 
systematics limited unlike the limits from 
CCD-based stacks saturate at a systematic floor of a few percent. 
The Resolve limits will therefore keep improving as the archive grows. 
Next-generation missions such as NewAthena and eXTP, with substantially 
larger effective areas, are forecast to probe a large fraction of 
the remaining parameter space of sterile neutrino 
DM~\citep{2016PhRvD..93f3518N,2020PhRvD.101l3009M}.

\myparagraph{Acknowledgments.}
The authors acknowledge support by the state of Baden-W\"urttemberg through bwHPC.  
This project has received funding through the MSCA4Ukraine project, grant number Ref 1.4 - UKR - 1245772 - MSCA4Ukraine, which is funded by the European Union. Views and opinions expressed are however those of the author(s) only and do not necessarily reflect those of the European Union, the European Research Executive Agency or the MSCA4Ukraine Consortium. Neither the European Union nor the European Research Executive Agency, nor the MSCA4Ukraine Consortium as a whole nor any individual member institutions of the MSCA4Ukraine Consortium can be held responsible for them. This work was partially supported by the Ministry of Education and Science of Ukraine under the project “Search for dark matter and particles beyond the Standard Model” (0125U002260). Supported by the Centre for Collective Use of Scientific Equipment “Laboratory of High Energy Physics and Astrophysics”, Taras Shevchenko National University of Kyiv.

\clearpage
\onecolumngrid
\setcounter{smsection}{0}

\section{Supplemental Material}
This Supplemental Material provides the technical details behind the
Letter, in the order in which the analysis uses them.
Section~\ref{sm:dataset} describes the dataset: the selection of the
XRISM/Resolve observations and the extraction of the spectra.
Section~\ref{sm:dm} gives the expected decay signal: the line flux,
the Milky Way halo model with its uncertainty, and the Doppler shift
of the line in each observation. Section~\ref{sm:fits} defines the
joint and the stacked fit and presents the limits as functions of the
line energy. Section~\ref{sm:stacked} examines the stacked spectrum
and the residual at $3.531$~keV, and Sec.~\ref{sm:origin} asks where
on the sky its photons come from. Tables~\ref{tab:obs}
and~\ref{tab:dm_params} at the end of the document list every
observation with its geometry, exposure, Doppler shift, column
density, and expected signal.


\smsection{The XRISM/Resolve dataset}{sm:dataset}

\smsubsection{Selection of the observations}{sm:ranking}

We start from all XRISM/Resolve observations that were publicly
available in April 2026~\citep{XRISM_DARTS_Public} and keep every
pointing whose cleaned event list has ONTIME $T > 1000$~s. We discard
the fields within $10^\circ$ of the Galactic Center, where the dark
matter (DM) density of the Milky Way is least certain. The remaining
observations are ranked by the total count rate $R$ of the archival
cleaned Resolve event list: in the low background regime of the Letter a line of
given flux is easiest to detect in the pointing with the lowest
background, so the ranked list runs from the most to the least
favorable observation. We select the $N_{\rm obs} = 60$
observations with the lowest count rates (excluding two pointings
with anomalous detector pixels, ObsIDs 000110000 and 000117000), up to
$R = 0.1352~\mathrm{cts\,s^{-1}}$. These 60 observations form the
sample used throughout the Letter. Table~\ref{tab:obs}
(p.~\pageref{tab:obs}) lists them; Table~\ref{tab:dm_params} gives
the Doppler shift, the column density, and the expected signal of
each pointing.

Among these 60 observations, nine are named RESOLVE\_THERM, with
ObsIDs in the 00xxxxxxx series that XRISM reserves for calibration
and engineering. They are short ($\simeq 1$~ks each), have some of
the lowest count rates of the archive, and contain no photons in the
energy band of our analysis. The expected signal in each of them is
below $0.01$~photons, and together they contribute $0.3\%$ of the
exposure of the sample. We therefore keep them rather than discarding
them; nothing below depends on this choice.

\smsubsection{Spectral extraction, aperture, and instrument response}{sm:aperture}

We extract the spectrum of every observation from the whole
$3' \times 3'$ Resolve field of view, following the standard XRISM
data reduction procedure~\citep{XRISM_ABC_Guide,Doyle2022_pipeline}.
At the count rates of our sample the grade branching ratios and the
spectral resolution are unaffected~\citep{xrism_pog}. Around
$3.5$~keV the Resolve line spread function (LSF) has a full width at
half maximum (FWHM) of $\simeq 5$~eV~\citep{xrism_pog}, and the native
channels are $0.5$~eV wide, so one resolution element spans about ten
channels. The effective area at this energy, read from the ancillary
response files (ARF), is $A_{\mathrm{eff}} \simeq 65~\mathrm{cm^2}$,
between $63$ and $67~\mathrm{cm^2}$ across the sample.

The decay signal is diffuse emission that fills the field of view. We
model it as a flat disc of radius $\vartheta_{\max} = 1.8'$, the
standard $78\%$ encircled energy extraction radius, and generate the
ARF of every observation for the same disc (\texttt{xaarfgen} with
\texttt{sourcetype=FLATCIRCLE}), so that the model flux and the
response refer to one and the same aperture, of solid angle
\begin{equation}
    \Delta\Omega
    = \pi\,\vartheta_{\max}^2
    = 8.613\times10^{-7}~\mathrm{sr},
    \label{eq:DeltaOmega}
\end{equation}
comparable to the field of view (the consistency of the fluxes is
ensured by the ARF, which is computed for the same aperture). All
fluxes quoted in the Letter
and below refer to this aperture. The response of an extended source
carries systematic uncertainties of instrumental origin: the
calibration, the statistical noise of the ray tracing that produces
the ARF, and the assumptions of the response generation
software~\citep{xrism_pog}. For our observations and band we estimate
their combined effect on the fluxes and limits at below one percent,
well below the uncertainty of the halo model
(Sec.~\ref{sm:halosys}).

\smsection{Expected decay signal}{sm:dm}

\smsubsection{Line flux and photon count}{sm:flux}

In this section we relate the parameters of the decaying DM particle
to the line flux and to the number of line photons expected in our
observations. A radiatively decaying DM particle of mass $m_{\dm}$
produces a monochromatic line at $E_\gamma = m_{\dm}/2$. For a fixed
lifetime the expected line flux depends on the DM column density
along the line of sight (los),
\begin{equation}
    S_{\dm}(\psi)  =  \int\limits_{\mathrm{los}} \rho\bigl(r(s,\psi)\bigr)\, ds,
    \label{eq:Sdm_def}
\end{equation}
where $\psi$ is the angle between the line of sight and the direction
towards the Galactic Center, $\rho(r)$ is the DM mass density
profile, $s$ is the distance along the los, and
\begin{equation}
    r(s,\psi) = \sqrt{R_\odot^2 + s^2 - 2R_\odot s\cos\psi}
    \label{eq:r_of_s}
\end{equation}
is the Galactocentric radius of a point at distance $s$. We adopt
$R_\odot = 8.2$~kpc for the distance from the Sun to the Galactic
Center~\cite{2016ARA&A..54..529B}.

The differential photon flux per unit solid angle in the direction
$\psi$ is
\begin{equation}
    \frac{d\Phi_\gamma}{d\Omega}(\psi)
     =
    \frac{\Gamma}{4\pi\, m_{\dm}}\, S_{\dm}(\psi),
    \label{eq:flux_general}
\end{equation}
where $\Gamma$ is the radiative decay rate. For a sterile neutrino of
mass $m_{\dm}$ and mixing angle $\theta$ it
is~\citep{1982PhRvD..25..766P,Barger:1995ty}
\begin{equation}
\Gamma \simeq
 1.361 \times 10^{-32} \left(\frac{\sin^2(2\theta)}{10^{-10}}\right)
 \left(\frac{m_{\dm}}{1~\mathrm{keV}}\right)^5~\mathrm{s}^{-1}.
 \label{eq:decay_rate}
\end{equation}
The flux collected within an aperture of solid angle
$\Delta\Omega$ is
\begin{equation}
    F_\gamma
     =
    \int\limits_{\Delta\Omega} \frac{d\Phi_\gamma}{d\Omega}(\psi)\, d\Omega
    \approx
    \frac{\Gamma}{4\pi\, m_{\dm}}\,
    S_{\dm}(\psi)\, \Delta\Omega,
    \label{eq:F_integrated}
\end{equation}
where we assumed that $S_{\dm}$ is approximately constant across the
field of view. For the aperture of Sec.~\ref{sm:aperture},
$\Delta\Omega = 8.613\times10^{-7}$~sr [Eq.~\eqref{eq:DeltaOmega}],
and substituting Eq.~\eqref{eq:decay_rate} the flux integrated over
the aperture becomes
\begin{equation}
F_{\gamma}
\approx
2.240 \times 10^{-8}
\left(
\frac{\sin^{2}(2\theta)}{10^{-10}}
\right)
\left(
\frac{m_{\dm}}{7~\mathrm{keV}}
\right)^{4}
\left(
\frac{S_{\dm}}
{10^{22}~\mathrm{GeV\,cm^{-2}}}
\right)
\left(
\frac{\Delta\Omega}
{8.613\times10^{-7}~\mathrm{sr}}
\right)
~\mathrm{ph\,cm^{-2}\,s^{-1}}.
    \label{eq:F_scaling}
\end{equation}

For an instrument with effective area $A_{\mathrm{eff}}$ and exposure
time $t_{\mathrm{obs}}$, the expected number of signal photons is
\begin{equation}
    N_\gamma
     =
    A_{\mathrm{eff}}\, t_{\mathrm{obs}}\, F_\gamma
     \simeq
    A_{\mathrm{eff}}\, t_{\mathrm{obs}}\,
    \frac{\Gamma}{4\pi\, m_{\dm}}\,
    S_{\dm}(\psi)\, \Delta\Omega.
    \label{eq:N_gamma}
\end{equation}
With the Resolve effective area at $3.5$~keV,
$A_{\mathrm{eff}} \simeq 65~\mathrm{cm^2}$ (Sec.~\ref{sm:aperture}),
Eq.~\eqref{eq:F_scaling} gives
\begin{equation}
    N_\gamma
    \approx
    2.9
    \left(\frac{\sin^2(2\theta)}{10^{-10}}\right)
    \left(\frac{m_{\dm}}{7~\mathrm{keV}}\right)^{4}
    \left(\frac{S_{\dm}}{10^{22}~\mathrm{GeV\,cm^{-2}}}\right)
    \left(\frac{A_{\mathrm{eff}}}{65~\mathrm{cm^2}}\right)
    \left(\frac{t_{\mathrm{obs}}}{2~\mathrm{Ms}}\right)
    \left(\frac{\Delta\Omega}{8.613\times10^{-7}~\mathrm{sr}}\right).
    \label{eq:N_gamma_scaling}
\end{equation}
At the benchmark $\sin^2(2\theta) = 2\times10^{-10}$, $m_{\dm} = 7$~keV,
and the sample column density $S_{\dm,\mathrm{eff}} = 2.13\times10^{22}~\mathrm{GeV\,cm^{-2}}$ [Eq.~\eqref{eq:Seff}]
this gives $N_\gamma \simeq 18$ line photons in a $2.9$~Ms exposure, consistent
with the sample total $\sum_i N_{\gamma,i} = 18.1$ (Sec.~\ref{sm:halo}).

\smsubsection{Milky Way dark matter halo model}{sm:halo}

To estimate the DM column density~\eqref{eq:Sdm_def} one needs the
dark matter profile of the Milky Way halo. We adopt the
Navarro--Frenk--White (NFW) profile~\citep{1996ApJ...462..563N}.
Rather than adopting one particular published mass model, we take a
benchmark value for its scale radius and fix its overall normalization
by requiring the local density to equal
$\rho_\odot = 0.4~\mathrm{GeV\,cm^{-3}}$~\citep{2010JCAP...08..004C,2011MNRAS.414.2446M,PDG2025}
at the Solar Galactocentric radius $R_\odot$: the scale radius of the
Milky Way halo is known only up to a factor of order two, so a generic
value is no less reliable than any single fit, while fixing
$\rho_\odot$ ties the profile to the one quantity that is directly
measured. At fixed shape the column density is proportional to
$\rho_\odot$, so a different choice of the local density simply scales
$S_{\dm}$, and with it the flux and the mixing angle limits, by the
same factor. The spread of published $\rho_\odot$ and the effect of
the halo shape on our limits are quantified in Sec.~\ref{sm:halosys}.

We truncate the integral in~\eqref{eq:Sdm_def} at the finite extent
of the halo, $r_{\mathrm{vir}} = 200~\mathrm{kpc}$, consistent with the
measured virial radius of the Milky Way,
$R_{200} = 218^{+12}_{-18}$~kpc~\citep{2020MNRAS.494.4291C}; the
neglected $r > r_{\mathrm{vir}}$ tail contributes $\lesssim 0.1\%$
of the column density. The cusp
of the NFW profile at the origin never enters the integral: the
pointings lie at angles $\psi > 38^\circ$ from the Galactic Center,
so no line of sight comes closer to the Galactic Center than
$R_\odot \sin\psi \simeq 5.1~\mathrm{kpc}$.

The NFW profile is
\begin{equation}
    \rho_{\mathrm{NFW}}(r)
     = 
    \frac{\rho_s}{(r/r_s)\,(1 + r/r_s)^2},
    \label{eq:NFW_profile}
\end{equation}
with the scale radius fixed to the benchmark value
$r_s = 20~\mathrm{kpc}$,
in the middle of the $r_s \approx 10$--$30~\mathrm{kpc}$ range allowed by
rotation curve fits~\citep{2017MNRAS.465...76M,2013JCAP...07..016N}. Fixing
$\rho(R_\odot) = \rho_\odot$ then sets the scale density,
\begin{equation}
    \rho_s = \rho_\odot\,x_\odot\,(1 + x_\odot)^2
    = 0.326~\mathrm{GeV\,cm^{-3}},
    \qquad x_\odot \equiv R_\odot/r_s.
\end{equation}

The line-of-sight integral in Eq.~\eqref{eq:Sdm_def} is evaluated
over $0 \le s \le s_{\max}(\psi)$, where
$s_{\max}(\psi) = R_\odot\cos\psi + \sqrt{r_{\rm vir}^2 - R_\odot^2\sin^2\psi}$
is the distance at which the line of sight exits the virial sphere of
radius $r_{\rm vir}$.

The selected pointings span
$\psi = 38$--$172^\circ$, all avoiding the inner Galaxy. The column densities
along these sight lines have two properties that limit halo
diagnostics. First, halo profiles normalized to the same $\rho_\odot$
at $R_\odot$ differ only in the inner few kiloparsecs, which these
sight lines never cross, so different halo shapes predict nearly the
same column density (Sec.~\ref{sm:halosys}). Second, the column density changes little
across the sample: it runs from
$\simeq 1.1\times10^{22}~\mathrm{GeV\,cm^{-2}}$ at the anticenter to
$\simeq 3.7\times10^{22}~\mathrm{GeV\,cm^{-2}}$ at the smallest
$\psi$, so a median split separates the two halves by only a
factor of $\sim\,$1.5. The factor $\sim\,$3--10
contrast between cored and cuspy haloes appears only toward the inner
Galaxy, which the present low count rate sample does not reach. The
column density diagnostics of the $3.531$~keV residual
(Sec.~\ref{sm:origin}) are therefore limited by
the geometry of the sample, not by its photon count.

The exposure weighted column density of the sample and the
corresponding $D$ factor are
\begin{equation}
    S_{\dm,\mathrm{eff}} = 2.128\times10^{22}~\mathrm{GeV\,cm^{-2}},
    \qquad
    D_{\mathrm{ap,eff}}
    = S_{\dm,\mathrm{eff}}\,\Delta\Omega
    = 1.833\times10^{16}~\mathrm{GeV\,cm^{-2}\,sr}.
    \label{eq:Seff}
\end{equation}
For $m_{\dm} = 7$~keV and $\sin^2(2\theta) = 2\times10^{-10}$ the
expected signal count of the sample, summed over the 60 pointings of
Table~\ref{tab:dm_params}, is $N_{\gamma,\mathrm{tot}} = 18.073$, and
the flux from the column density~\eqref{eq:Seff} is
$F_{\gamma,\mathrm{eff}} = 9.533\times10^{-8}~\mathrm{ph\,cm^{-2}\,s^{-1}}$.

\smsubsection{Uncertainty of the halo model}{sm:halosys}

\begin{figure}[t]
\centering
\includegraphics[width=\linewidth]{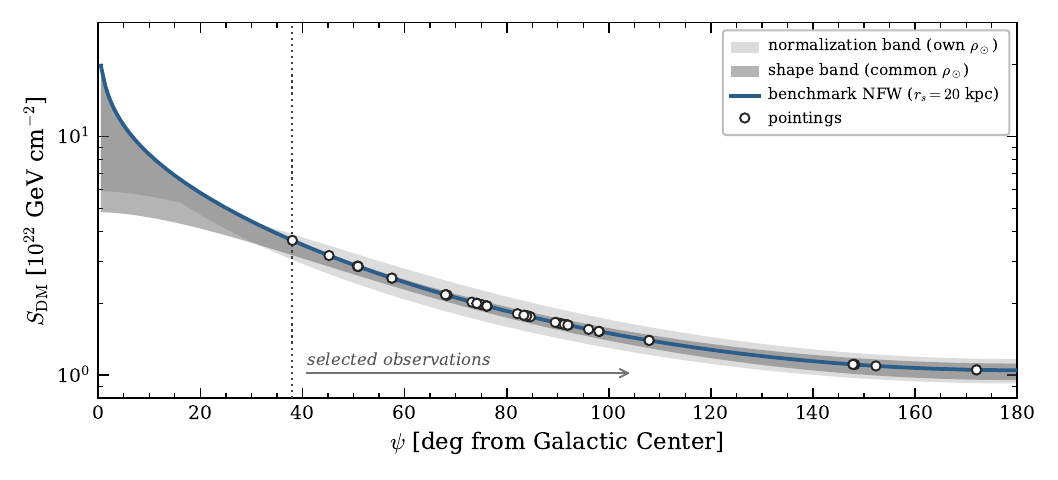}
\caption{Dark matter column density $S_{\dm}(\psi)$ as a function of
the angle $\psi$ to the Galactic Center. Solid line: the baseline NFW
profile ($\rho_\odot = 0.4~\mathrm{GeV\,cm^{-3}}$ at
$R_\odot = 8.2$~kpc), drawn for $\psi > 38^\circ$ (vertical line),
the range of the sample; open circles mark the 60 pointings.
The lighter band spans the Milky Way mass models of
Table~\ref{tab:massmodels}, each with its own $\rho_\odot$; the
darker band spans the same models rescaled to
$\rho_\odot = 0.4~\mathrm{GeV\,cm^{-3}}$ and is about $13\%$ wide at
a given angle. Toward the inner Galaxy ($\psi \lesssim 38^\circ$),
not sampled by our observations, the bands widen as cored and cuspy
profiles diverge.}
\label{fig:sdm_psi}
\end{figure}

Here we quantify how the choice of the halo model affects
$S_{\dm,\mathrm{eff}}$ and, through it, the limits. The column
density, and with it the flux and mixing angle limits, scales with
the halo normalization. Determinations of the local density span
$\rho_\odot \simeq 0.3$--$0.6~\mathrm{GeV\,cm^{-3}}$~\citep{2021PDU....3200826B,2021RPPh...84j4901D},
with individual measurements clustering around the value adopted
here: $0.389 \pm 0.025$~\citep{2010JCAP...08..004C},
$0.40 \pm 0.04$~\citep{2011MNRAS.414.2446M}, and the Particle Data
Group average
$0.39\pm0.03~\mathrm{GeV\,cm^{-3}}$~\citep{PDG2025}. Rescaling
$\rho_\odot$ alone over this range would change the column density
by a factor of $\sim\,$2, but the rotation curve ties $\rho_\odot$
and $r_s$ together: a model with a lower local density comes with a
larger scale radius, and the two shifts partly cancel along the outer
sight lines of this sample. We therefore recompute
$S_{\dm,\mathrm{eff}}$ under a set of self-consistent Milky Way mass
models, each with its own shape and its own $(\rho_\odot, r_s)$ pair
(Table~\ref{tab:massmodels}). The models span the cored (Burkert) and
cuspy (NFW) profile families; the Burkert entry is the rotation curve
fit of Nesti \& Salucci~\citep{2013JCAP...07..016N}, whose cored shape
prefers a higher local density,
$\rho_\odot = 0.487~\mathrm{GeV\,cm^{-3}}$. The models give
$S_{\dm,\mathrm{eff}}$ between $1.81$ and
$2.40\times10^{22}~\mathrm{GeV\,cm^{-2}}$ (a factor of 1.32 rather
than 2), bracketing the baseline value of Eq.~\eqref{eq:Seff}.
Relative to the baseline the column density shifts by $-15\%$ under
the most conservative model (the contracted NFW fit
of~\citep{2020MNRAS.494.4291C}) and by $+13\%$ under the cored
Burkert fit of Nesti \& Salucci~\citep{2013JCAP...07..016N}. Since
the flux and mixing angle limits scale as $1/S_{\dm,\mathrm{eff}}$,
they correspondingly weaken by at most $\sim 18\%$ and tighten by at
most $\sim 11\%$. Rescaling every model of
Table~\ref{tab:massmodels} to the common
$\rho_\odot = 0.4~\mathrm{GeV\,cm^{-3}}$ isolates the effect of the
halo shape: $S_{\dm,\mathrm{eff}}$ then changes by between $-8.1\%$
and $+3.1\%$ relative to the baseline, the largest shift coming from
the cored Burkert fit. Figure~\ref{fig:sdm_psi} shows both effects
as functions of the angle to the Galactic Center.

\begin{table}[t]
\centering
\caption{Exposure weighted dark matter column density of the sample
under self-consistent Milky Way mass models, each with its own local
density $\rho_\odot$ and scale radius $r_s$. The last column is the
ratio to the baseline normalization
($\rho_\odot = 0.4~\mathrm{GeV\,cm^{-3}}$, NFW $r_s = 20$~kpc,
$R_\odot = 8.2$~kpc). For the Nesti \& Salucci
model~\citep{2013JCAP...07..016N} the profile is a Burkert form, the
$r_s$ column quotes its core radius $r_c$, and their fitted
$R_\odot = 7.94$~kpc is used.}
\label{tab:massmodels}
\setlength{\tabcolsep}{6pt}
\begin{tabular}{llccc}
\toprule
Model & Profile & $\rho_\odot$ & $r_s$ & $S_{\dm,\mathrm{eff}}/S^{\rm base}_{\dm,\mathrm{eff}}$ \\
 & & (GeV\,cm$^{-3}$) & (kpc) & \\
\midrule
Baseline (this work) & NFW & 0.40 & 20.0 & 1.00 \\
Nesti 2013~\citep{2013JCAP...07..016N} & Burkert & 0.49 & 9.3 & 1.13 \\
McMillan 2017~\citep{2017MNRAS.465...76M} & NFW & 0.38 & 19.6 & 0.95 \\
Cautun 2020~\citep{2020MNRAS.494.4291C} & contr.\ NFW & 0.33 & 23.8 & 0.85 \\
PDG 2025~\citep{PDG2025} & NFW & 0.39 & 20.0 & 0.97 \\
\bottomrule
\end{tabular}
\end{table}

So far the halo has been modeled as spherical. To quantify the effect
of a possible flattening, we replace $r \to \sqrt{R^2 + z^2/q^2}$ in
Eq.~\eqref{eq:NFW_profile}, keep $\rho_\odot$ fixed at the Solar
position, and recompute $S_{\dm,\mathrm{eff}}$. For the axis ratios
$q = 0.9$--$1.1$ spanned by recent determinations of the Milky Way
halo shape~\citep{2016ApJ...833...31B,2019MNRAS.485.3296W},
$S_{\dm,\mathrm{eff}}$ changes by at most $\pm 5\%$; even an extreme
oblateness $q = 0.8$ lowers it by only $11\%$. Flattening at this
level is subdominant to the normalization uncertainty.

\smsubsection{Per observation Doppler correction}{sm:doppler}

Unlike in the previous, lower resolution searches with CCD-based instruments, at the Resolve energy resolution ($\mathrm{FWHM} \simeq 5$~eV)
the motion of the observer with respect to the Galactic rest frame is
no longer negligible: it shifts the expected line energy by a
fraction of the line spread function, and the shift
differs from one observation to the next. We therefore apply a per
observation Doppler correction, taking the observer motion to be the
sum of the solar and terrestrial contributions. We adopt the solar
peculiar velocity with respect to the Local Standard of Rest,
$\vec{V}_{\mathrm{pec}} = (U,V,W)_\odot = (11.10^{+0.69}_{-0.75}, 12.24 \pm 0.47, 7.25^{+0.37}_{-0.36})~\mathrm{km\,s^{-1}}$~\citep{2010MNRAS.403.1829S}
in the Galactic frame ($U$ toward the Galactic Center, $V$ in the
direction of Galactic rotation, $W$ toward the North Galactic Pole),
with additional systematic uncertainties
$\sim\,(1,\,2,\,0.5)~\mathrm{km\,s^{-1}}$~\citep{2010MNRAS.403.1829S}.
The total velocity is
$\vec{V}_\odot = V_{\mathrm{LSR}} (0,1,0) + \vec{V}_{\mathrm{pec}}$, 
where $V_{\mathrm{LSR}} = 229.0 \pm 0.2\,(\mathrm{stat.})~\mathrm{km\,s^{-1}}$ is the circular velocity at the solar radius, with a systematic uncertainty of $\lesssim 3\%$~\citep{2019ApJ...871..120E};
a recent independent determination gives
$234.0 \pm 1.4~\mathrm{km\,s^{-1}}$~\citep{2023ApJ...946...73Z}.
In addition, we include the orbital motion of the Earth
($|v_\oplus| \le 29.8~\mathrm{km\,s^{-1}}$, which leads to a shift
of up to $\pm0.35$~eV, or $\sim\,$0.7 Resolve PI channels, at
3.53~keV) projected onto each line of sight at the epoch of the
corresponding observation. The Earth's diurnal rotation
($\le 0.46~\mathrm{km\,s^{-1}}$) and the spacecraft's orbital motion
($\simeq 7.6~\mathrm{km\,s^{-1}}$ in low Earth orbit, averaging out
over exposures spanning many $\sim\,$96-minute orbits) are
negligible. The resulting line of sight velocities and observed line
energies of every observation are listed in Table~\ref{tab:dm_params};
across the sample the total shift spans $-2.7$ to $+2.9$~eV
at $3.5$~keV.

The uncertainties of the adopted velocities (the solar peculiar
velocity, $V_{\rm LSR}$ with its $\lesssim 3\%$ systematic, and the
spread among recent determinations) translate into at most
$\simeq 8~\mathrm{km\,s^{-1}}$ along any line of sight, i.e.\
$\lesssim 0.1$~eV at $3.5$~keV (about $2\%$ of the LSF FWHM).
Shifting all per observation line energies coherently by
$\pm 0.05$~eV ($\pm 0.1$~eV) changes the joint fit test statistic by
$|\Delta C| < 0.003$ ($0.006$) and the $2\sigma$ flux limit by less
than $0.1\%$. This systematic is negligible.

\smsection{Fitting procedure}{sm:fits}

In this work we derive the limits with two fitting procedures, the
\emph{joint fit} and the \emph{stacked fit} as described below.

Both fits use the $3.43$--$3.57$~keV band (the fitting band; 282 native $0.5$~eV
channels) and the C statistic. The power law represents the
featureless astrophysical and instrumental background; its photon
index is frozen at
$\Gamma_{\rm ph} = 1.46$~\citep{1997MNRAS.285..449C,2002ApJ...576..188M},
and varying $\Gamma_{\rm ph}$ over the range of plausible background
slopes changes the derived limits by less than $1\%$. The Gaussian
models the decay line, with a width $\sigma_E = 2$~eV set by the
virial broadening in the Galactic halo (Sec.~\ref{sm:search}
discusses this width further). The significance of a line is
$\sqrt{\Delta C}\sigma$, where $\Delta C$ is the increase of the C statistic
when the line normalization is set to zero and all other free
parameters are refitted. Upper limits are profile likelihood bounds
on the line normalization, $\Delta C = 4$ ($2\sigma$) and
$\Delta C = 9$ ($3\sigma$) above the best fit, with the flux
constrained to be non-negative. Quoted uncertainties are $1\sigma$
from the covariance matrix; asymmetric intervals are profile
likelihood ($\Delta C = 1$) ranges.

\textbf{Joint fit.} In the joint fit the spectra of all observations are fitted
simultaneously. 
Each observation has its own power law normalization $c_i$ and
its own line energy $E_{\mathrm{obs},i}$, Doppler shifted as
described in Sec.~\ref{sm:doppler}. The line normalizations are tied
through the column densities: for each pointing we compute the angle
$\psi_i$ between its center and the Galactic Center and the column
density $S_{\dm}(\psi_i)$ of the baseline model of the halo,
Sec.~\ref{sm:halo}, and tie the normalizations via the ratios $w_i$,
so that a single amplitude $A$ describes the whole sample. We quote
this amplitude for the reference pointing ObsID~000114000
($S_{\dm,\mathrm{ref}} = 1.948\times10^{22}~\mathrm{GeV\,cm^{-2}}$,
Table~\ref{tab:dm_params}); the choice of the reference is a
relabeling of the tied normalization and does not affect the limits.

\textbf{Stacked fit.}
In the stacked fit the sample is co-added into a single spectrum
before the fit. The counts are summed channel by channel,
with one continuum normalization and one line flux. 
The individual responses are exposure weighted averages of the ARF
and the RMF. 
When the spectra are stacked as recorded, we call the result the
\emph{detector frame} co-added spectrum. 
We also build a \emph{Doppler corrected} co-added spectrum (Sec.~\ref{sm:doppcorr}), in which
every spectrum is moved to the halo rest frame before the sum. 
For each observation $i$ the channels  are relabeled from their recorded energy
$E$ to $E - \Delta E_i$, with the shift $\Delta E_i$ of
Table~\ref{tab:dm_params} rounded to the nearest whole number of
$0.5$~eV channels (a rounding error below $0.25$~eV).
The folded models of each observation are relabeled by the same integer as its
counts, so the response is always applied at the recorded energy and
no averaged response is formed. 
The shifts run from $-5$ to $+6$
channels, so the band common to all shifted spectra is 11 channels
narrower than the fitting band, and the Doppler corrected co-added spectrum has 271 channels.

\begin{figure}[t]
\centering
\includegraphics[width=\linewidth]{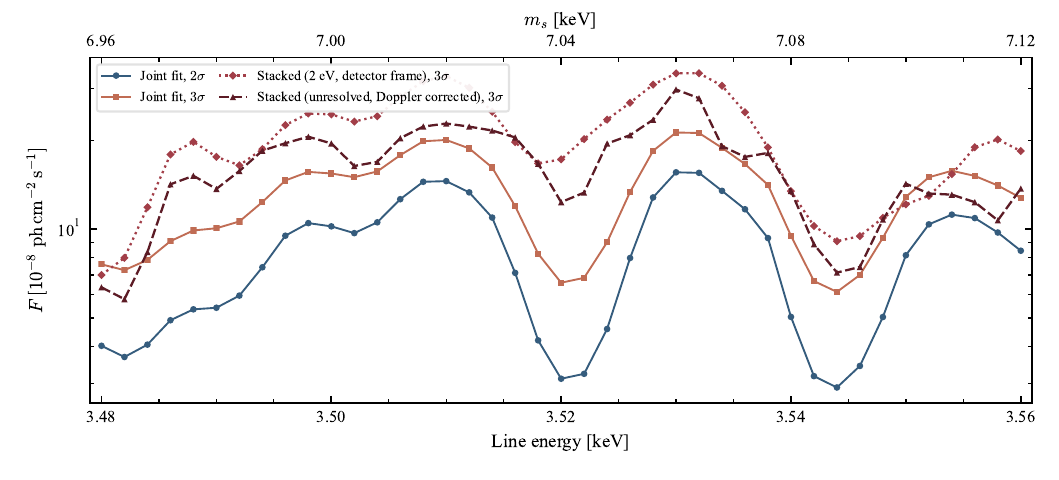}
\caption{Upper limits on the flux of a line as functions of
the line energy (\emph{bottom axis}) and of the sterile neutrino mass
$m_s = 2E$ (\emph{top axis}). Blue circles and orange squares show the
$2\sigma$ and $3\sigma$ joint fit limits, quoted, as in the text,
for the reference pointing (ObsID~000114000); red diamonds (dotted) show the $3\sigma$ limit of the stacked fit for the $\sigma_E = 2$~eV line in the detector frame co-added spectrum, and dark red triangles (dashed) the same limit for an unresolved line in the Doppler corrected co-added spectrum, both quoted for the exposure weighted average flux of the sample.}
\label{fig:flux_limits}
\end{figure}

\begin{figure}[t]
\centering
\includegraphics[width=\linewidth]{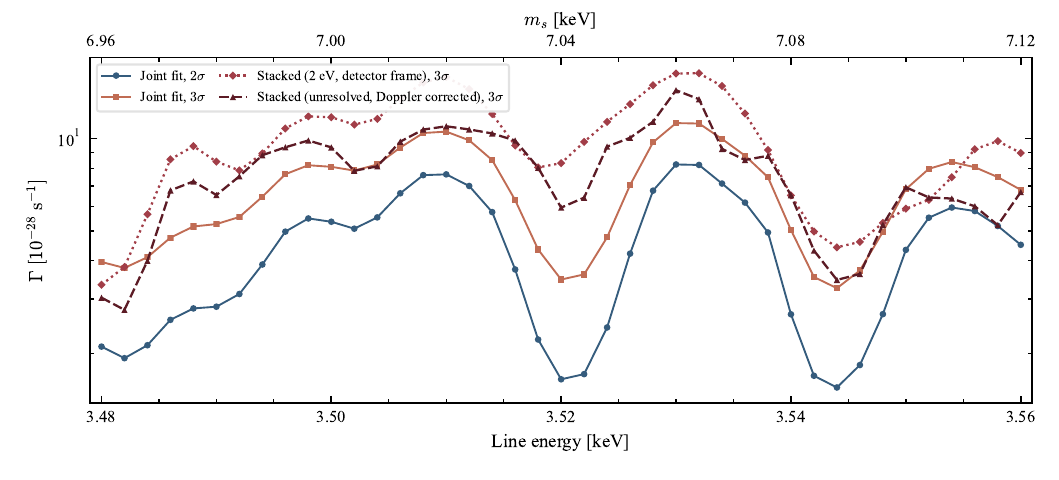}
\caption{Upper limits on the radiative decay width $\Gamma$ of the
DM candidate as functions of the line energy (\emph{bottom axis}) and of
the sterile neutrino mass $m_s = 2E$ (\emph{top axis}). Blue circles and
orange squares show the $2\sigma$ and $3\sigma$ joint fit limits;
red diamonds (dotted) and dark red triangles (dashed) show the $3\sigma$ stacked limits for the $\sigma_E = 2$~eV line in the detector frame co-added spectrum and for an unresolved line in the Doppler corrected co-added spectrum.
$\Gamma$ is computed at each energy from the corresponding
mixing angle limit and does not depend on the flux convention.}
\label{fig:gamma_limits}
\end{figure}

Figures~\ref{fig:flux_limits} and~\ref{fig:gamma_limits}
show the limits as functions of the line energy.
The limits are broadly consistent.
Figure~\ref{fig:flux_limits} gives the limits on the line flux, each
fit in its own convention: the joint fit limit
is the flux of the reference pointing, and the stacked limit the
exposure weighted average flux of the sample.
Figure~\ref{fig:gamma_limits} recasts the same limits as the
radiative decay width $\Gamma$, computed at each energy from the
corresponding mixing angle limit through Eq.~\eqref{eq:decay_rate}.
Figure~1 of the Letter shows the same limits in
terms of the mixing angle.

\smsection{Stacked spectrum and the $3.531$~keV residual}{sm:stacked}

In this section we examine the stacked spectrum of the sample for
residuals. Table~\ref{tab:residual} summarizes the outcome, the rest of the section supplies the details. We first describe the fitting procedure and
the known atomic lines in the band (Sec.~\ref{sm:fitsetup}),
then search for an additional line in the
detector frame co-added spectrum (Sec.~\ref{sm:search}) and in the Doppler corrected
co-added spectrum (Sec.~\ref{sm:doppcorr}), and test the instrumental and atomic counterparts of the residuals we find (Sec.~\ref{sm:counterparts}).
Section~\ref{sm:origin} then asks where on the sky the photons of
the residual come from.

\begin{table*}[t]
\centering
\caption{Significance of the $3.531$~keV residual. 
Each row reports the increase of the C statistic when the 3.531~keV line is present in the model.
The line width is either $\sigma_E = 2$~eV as expected for a halo line or an unresolved width
($\sigma_E = 10^{-3}$~eV). 
The continuum model is either a powerlaw or the powerlaw plus the
\ion{Cl}{XVII}/\ion{K}{XVIII} complex (``Cl/K'') at $3.5089$~keV. 
The last
column gives $\Delta C$ of the Cl/K itself, fitted alone on the
continuum. 
For every configuration the table gives $\Delta C$ with the local significance in
parentheses, in units of $\sigma$, and the best fit line flux $F$ in units of
$10^{-8}$~ph\,cm$^{-2}$\,s$^{-1}$ with $\Delta C = 1$ errors;
the joint fit flux refers to the reference pointing.
The
global significance is derived in Sec.~\ref{sm:global}.}
\label{tab:residual}
\setlength{\tabcolsep}{3.5pt}
\footnotesize
\resizebox{\textwidth}{!}{%
\begin{tabular}{l cc cc cc cc cc}
\toprule
 & \multicolumn{4}{c}{$\sigma_E = 2$~eV line} & \multicolumn{4}{c}{unresolved line} & \multicolumn{2}{c}{Cl/K} \\
\cmidrule(lr){2-5}\cmidrule(lr){6-9}
 & \multicolumn{2}{c}{continuum} & \multicolumn{2}{c}{continuum $+$ Cl/K} & \multicolumn{2}{c}{continuum} & \multicolumn{2}{c}{continuum $+$ Cl/K} & \multicolumn{2}{c}{alone} \\
\cmidrule(lr){2-3}\cmidrule(lr){4-5}\cmidrule(lr){6-7}\cmidrule(lr){8-9}\cmidrule(lr){10-11}
 & $\Delta C$ ($\sigma$) & $F$ & $\Delta C$ ($\sigma$) & $F$ & $\Delta C$ ($\sigma$) & $F$ & $\Delta C$ ($\sigma$) & $F$ & $\Delta C$ ($\sigma$) & $F$ \\
\midrule
detector frame    & 6.37 (2.5) & $14.1^{+6.3}_{-5.9}$ & 7.12 (2.7) & $14.9^{+6.3}_{-5.9}$ & 6.96 (2.6) & $12.0^{+5.2}_{-4.8}$ & 7.58 (2.8) & $12.4^{+5.2}_{-4.8}$ & 4.76 (2.2) & $10.0^{+5.2}_{-4.8}$ \\
Doppler corrected & 6.67 (2.6) & $14.4^{+6.3}_{-5.9}$ & 7.17 (2.7) & $14.9^{+6.2}_{-5.9}$ & 8.36 (2.9) & $13.1^{+5.3}_{-4.9}$ & 8.81 (3.0) & $13.5^{+5.3}_{-4.9}$ & 1.76 (1.3) & $6.1^{+5.1}_{-4.7}$ \\
hybrid            & 6.63 (2.6) & $14.3^{+6.2}_{-5.9}$ & 7.45 (2.7) & $14.7^{+6.8}_{-5.3}$ & 8.29 (2.9) & $13.1^{+5.3}_{-4.9}$ & 9.03 (3.0) & $13.5^{+5.5}_{-4.6}$ & 4.58 (2.1) & $9.8^{+5.2}_{-4.8}$ \\
joint fit         & 2.11 (1.5) & $5.9^{+4.7}_{-4.2}$ & 2.51 (1.6) & $6.5^{+4.7}_{-4.2}$ & 2.59 (1.6) & $5.6^{+4.2}_{-3.7}$ & 2.92 (1.7) & $6.0^{+4.2}_{-3.7}$ & 5.0 (2.2) & $7.8^{+4.3}_{-3.8}$ \\
\bottomrule
\end{tabular}}
\end{table*}

\smsubsection{Fitting the stacked spectrum}{sm:fitsetup}

We fit the stacked spectrum in the fitting band with the model of Sec.~\ref{sm:fits}: a
power law with $\Gamma_{\rm ph} = 1.46$ frozen plus
Gaussian lines representing the known atomic lines in the band and
the candidate line. The fitting band is deliberately broader than
the $3.48$--$3.56$~keV interval over which the limits and the
residual scan are reported, so that results near the ends of the
reported interval are unaffected by the boundaries of the fit
window. Quoted uncertainties are profile likelihood ($\Delta C = 1$)
ranges with the other parameters refitted. All significances in this section are
local: they do not account for the number of energies searched
across the band, which we take into account at the end of
Sec.~\ref{sm:global}.

Two known atomic complexes fall within the fitting
band~\citep{Foster:2012hy}. The first is formed by the
intercombination components of the \ion{K}{XVIII} He$\alpha$
multiplet, ($3.4930$~keV and $3.4965$~keV, emissivity weighted centroid $3.4946$~keV). 
The second is the \ion{Cl}{XVII} Ly$\beta$ doublet
($3.5077$ and $3.5088$~keV, emissivity weighted centroid
$3.5085$~keV), blended at Resolve resolution with the
\ion{K}{XVIII} resonance component at $3.5105$~keV, together
forming the \ion{Cl}{XVII}/\ion{K}{XVIII} complex, mentioned in the main
text.

\smsubsection{Search for an additional line}{sm:search}

We search for a putative DM line on top of two baseline models. The
\emph{continuum only} model is the bare power-law continuum. The
\emph{continuum plus complex} model adds the
\ion{Cl}{XVII}/\ion{K}{XVIII} complex described above, modeled
as a narrow Gaussian at the measured position of the complex
($3.5089$~keV) with its normalization left free.

To the continuum only model we add a single Gaussian. Its width is frozen at $\sigma_E = 2$~eV,
which matches the expected Doppler (virial) width of a Milky Way halo
DM line and encompasses the thermal and turbulent broadening of the
astrophysical lines (the purely thermal width of K and Cl ions at
their formation temperature is $\sigma_E \simeq 0.8$~eV). 
For the detector frame co-added spectrum, the Doppler shifts would further add $\simeq 1.7$~eV in
quadrature, giving an effective width of $\simeq 2.6$~eV.

\begin{figure*}[t]
\centering
\includegraphics[width=\textwidth]{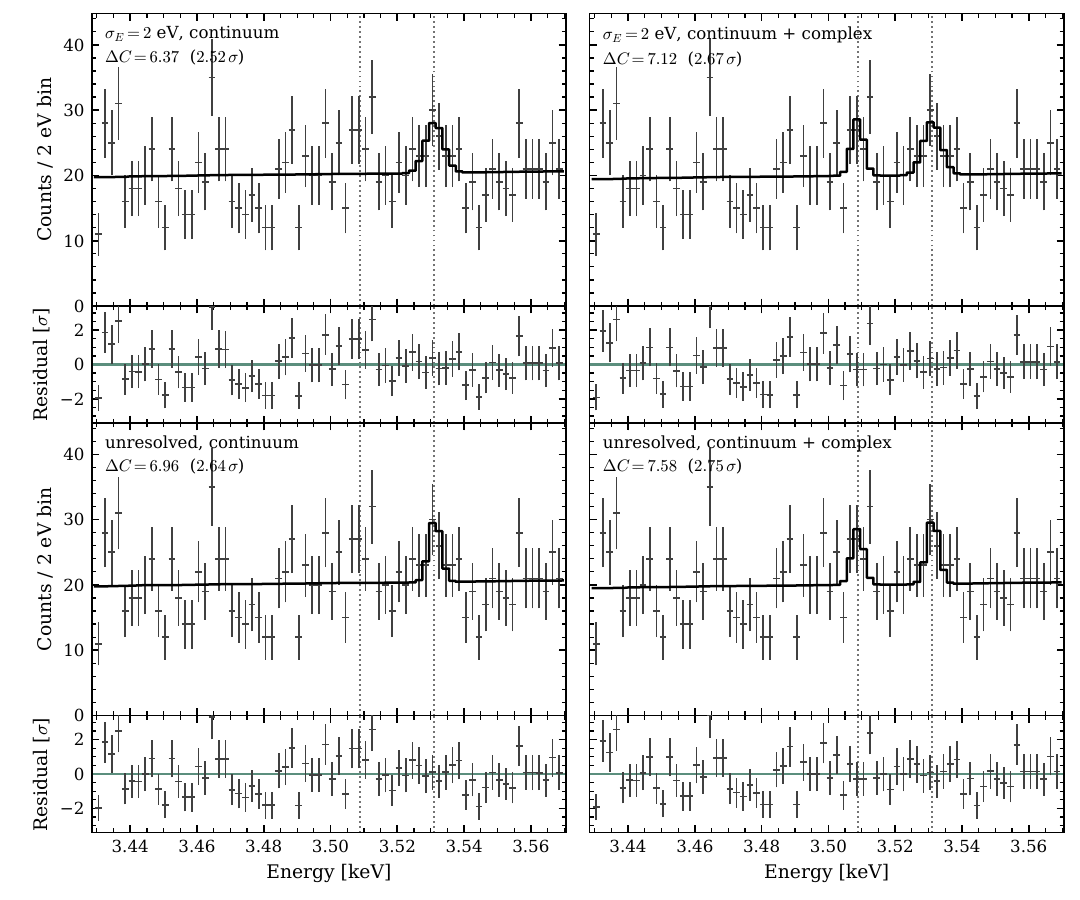}
\caption{Detector frame co-added spectrum of the 60 observations in the $3.43$--$3.57$~keV band, shown in 2-eV display bins with Poisson errors, with the best fit models (red). \emph{Top row:}
the DM line as a $\sigma_E = 2$~eV Gaussian at $3.5311$~keV;
\emph{bottom row:} as an unresolved line. \emph{Left column:} on the
continuum only model; \emph{right column:} on the continuum plus the
\ion{Cl}{XVII}/\ion{K}{XVIII} complex at $3.5089$~keV. Dotted
lines mark the two energies. Each panel gives $\Delta C$ and the
local significance of the DM line, and the strip below it the
residuals in units of $1\sigma$. The values are those of the first
row of Table~\ref{tab:residual}.}
\label{fig:spectrum_nominal}
\end{figure*}

The scan 
marginally detects the \ion{Cl}{XVII}/\ion{K}{XVIII} complex
at $3.5094 \pm 0.0018$~keV ($\Delta C = 4.95$, $2.2\sigma$,
$F = 12.6^{+6.3}_{-5.9}\times10^{-8}$~ph\,cm$^{-2}$\,s$^{-1}$) and finds an
excess at
$E = 3.5311^{+0.0010}_{-0.0013}$~keV ($\Delta C = 6.37$, $2.5\sigma$), see
Figure~\ref{fig:spectrum_nominal}. 
The latter excess is not associated with any known atomic line, see Sec.~\ref{sm:counterparts}.
Freeing the width of the $3.531$~keV excess yields a
marginal ($\Delta C = 0.59$) preference for an intrinsically narrow
line over the $\sigma_E = 2$~eV width; the width itself is only
weakly constrained, $\sigma_E < 8.5$~eV at the $2\sigma$ level. With
the unresolved width the excess has $\Delta C = 6.96$ ($2.6\sigma$).

On the continuum plus complex model, scanning the position of an
additional $\sigma_E = 2$~eV line again finds the excess at
$E = 3.5311$~keV, with a local significance $2.7\sigma$, and the unresolved width
raises it to $2.8\sigma$ ($\Delta C = 7.58$; right
column of Fig.~\ref{fig:spectrum_nominal}). The complex included in
the model is recovered at $2.2\sigma$, and
freezing it at its best fit value changes $\Delta C$ of the DM line
by less than $0.02$.

Both baseline models thus detect the $3.531$~keV residual at
$2.5$--$2.8\sigma$ local significance; the global significance is
derived in Sec.~\ref{sm:global}.

\smsubsection{Doppler corrected stacked spectrum}{sm:doppcorr}

\begin{figure*}[t]
\centering
\includegraphics[width=\textwidth]{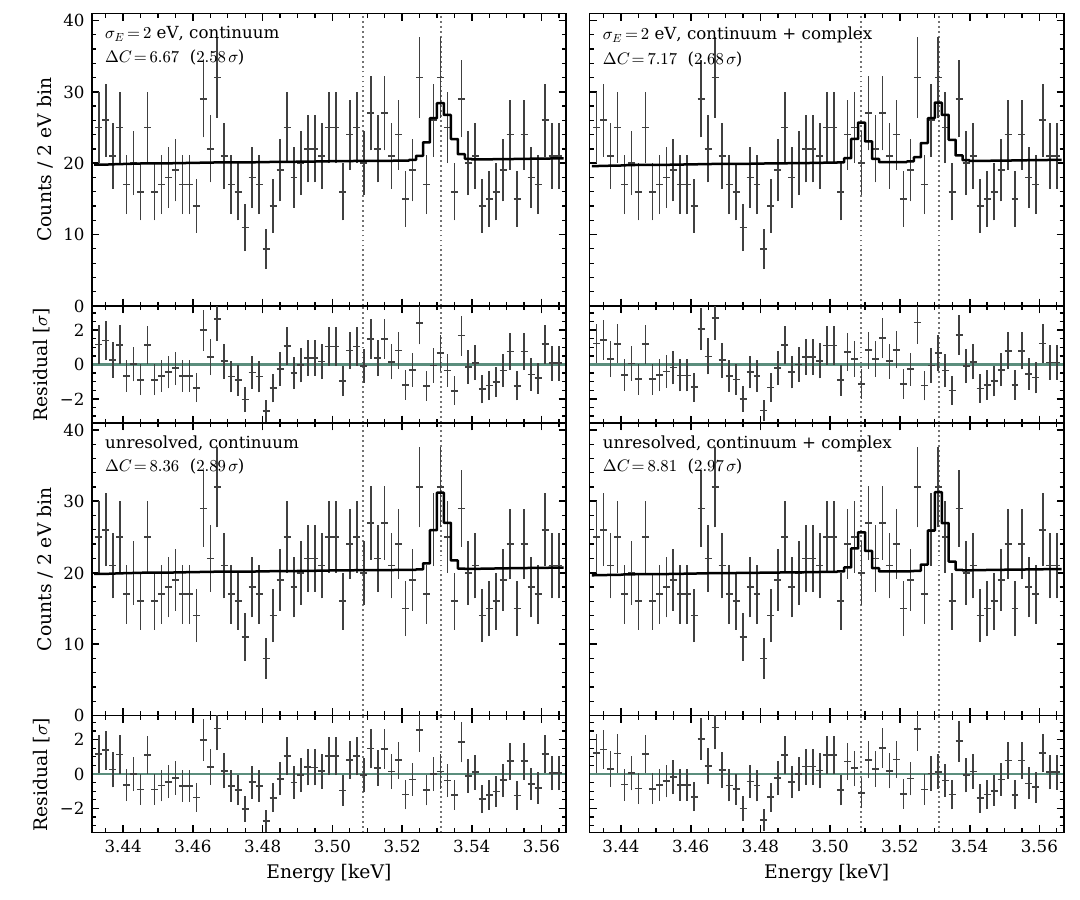}
\caption{Same as Fig.~\ref{fig:spectrum_nominal}, for the Doppler
corrected co-added spectrum: every spectrum is moved to the halo rest frame
before the sum, which leaves 271 channels. The values are those of
the second row of Table~\ref{tab:residual}.}
\label{fig:spectrum_doppler}
\end{figure*}
We also perform a search for unidentified line in the Doppler corrected stacked spectrum. 
The correction raises the significance of the residual, from $6.4$
to $6.7$ in $\Delta C$ for the $\sigma_E = 2$~eV line and from $7.0$
to $8.4$ for the unresolved line, that is from $2.5\sigma$ to $2.6\sigma$ and from $2.6\sigma$ to
$2.9\sigma$; with the complex in the model the unresolved line
reaches $\Delta C = 8.81$, $3.0\sigma$ local ($2.0$--$2.3\sigma$ global,
Sec.~\ref{sm:global}). 
With the complex in the model the unresolved line carries 26 photons (fluxes are collected in Table~\ref{tab:residual}), and a free energy fit returns $E = 3.5307^{+0.0007}_{-0.0008}$~keV with a width below $6.5$~eV at $2\sigma$. The gain comes from the observations with the largest
shifts, where the correction has something to correct.

\begin{figure*}[t]
\centering
\includegraphics[width=\textwidth]{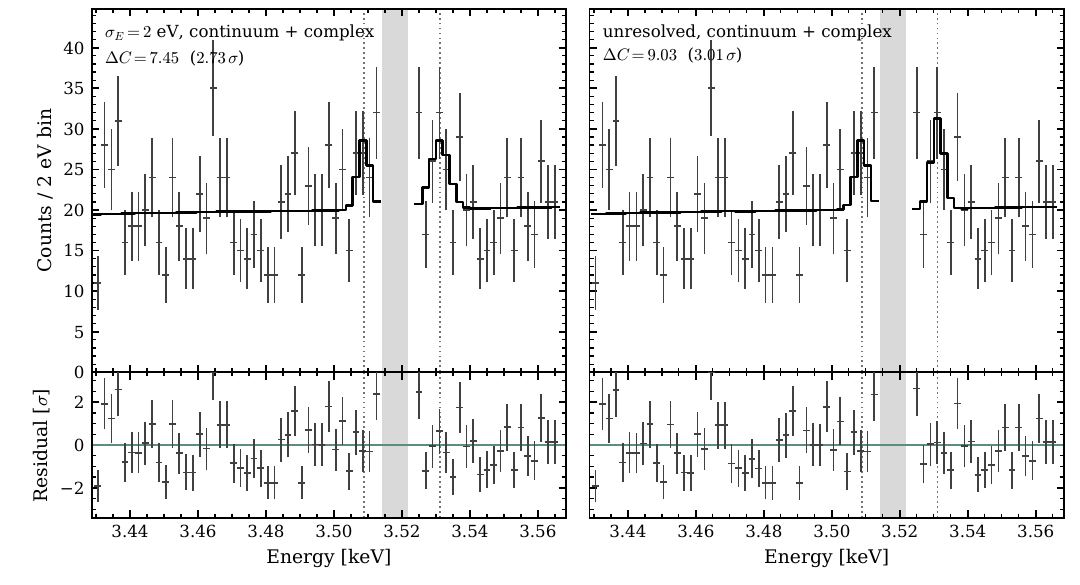}
\caption{Hybrid co-added spectrum of the 60 observations in 2-eV display bins
with Poisson errors, and the best fit continuum plus complex model
with the DM line as a $\sigma_E = 2$~eV Gaussian (\emph{left}) or
an unresolved line (\emph{right}). Channels below $3.5145$~keV are
co-added as recorded and channels above $3.5215$~keV after moving
each spectrum to the halo rest frame; the gray band is the guard
zone excluded from the fit. Dotted lines mark the complex and the
residual. Each panel gives $\Delta C$ and the local significance of
the DM line; the strips below give the residuals in units of
$1\sigma$. The fit uses 257 native channels.}
\label{fig:hybrid}
\end{figure*}

The \ion{Cl}{XVII}/\ion{K}{XVIII} complex provides a negative
control. A line that sits at a fixed energy in every observation is
smeared by the correction exactly as a halo line is sharpened, and
the $\Delta C$ of the complex drops from $4.8$ in the detector frame co-added spectrum to $1.8$ in the Doppler corrected one (last column of
Table~\ref{tab:residual}). The residual and the complex therefore
respond to the correction in opposite directions, as expected if the
first moves with the halo and the second does not.

This opposite response suggests a third construction, the \textit{hybrid}
co-added spectrum, in which the channels around the complex stay in the detector frame and
the channels around the residual are Doppler corrected. We split the
band at $3.518$~keV, midway between the two features, and exclude a
guard zone of 7 channels on either side of the split,
$3.5145$--$3.5215$~keV, so that no channel of any observation is used
twice (the guard of 14 channels exceeds the largest shift of 6). The
low piece, 170 channels, is co-added as recorded and carries the
complex at $3.5089$~keV; the high piece, 87 channels, is co-added
after the correction and carries the line at $3.5311$~keV; the two
pieces share one continuum normalization and one C statistic, and
each observation loses 25 of its 282 channels to the guard and the
trim. With no correction and no guard the hybrid co-added spectrum reduces to
the detector frame co-added one, and with every channel corrected and no guard to the Doppler corrected one. Figure~\ref{fig:hybrid} shows the hybrid
spectrum. The line comes out as in the Doppler corrected co-added spectrum and
the complex as in the detector frame one (third row of
Table~\ref{tab:residual}): with the complex in the model the
residual has $\Delta C = 7.45$ for the $2$~eV line and $9.03$ for the unresolved line ($3.0\sigma$ local), and the complex has $\Delta C = 4.58$. Scanning the energies in the hybrid
co-added spectrum returns $3.5305$~keV for the line and $3.5085$~keV for the
complex, within one channel of the fixed values.

\begin{figure}[t]
\centering
\includegraphics[width=\linewidth]{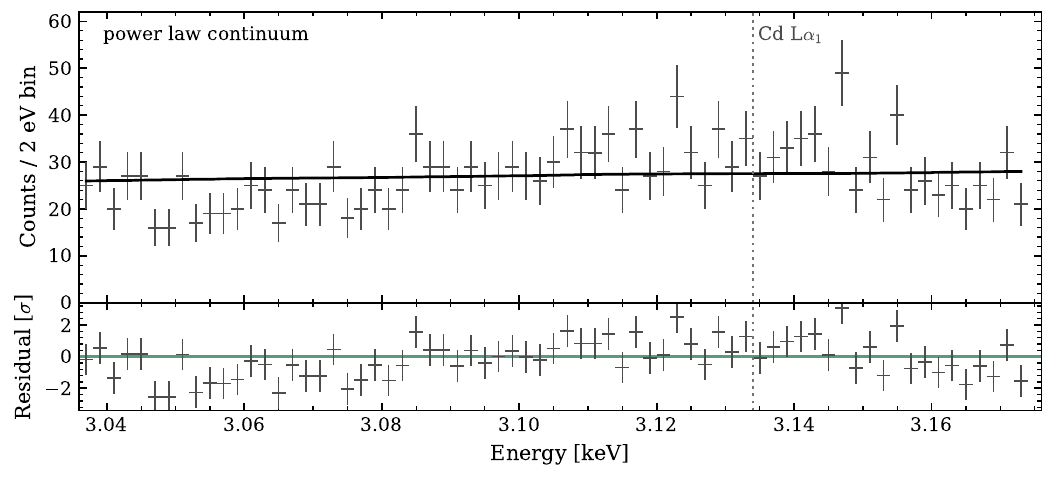}
\caption{Stacked XRISM/Resolve spectrum in the $3.036$--$3.176$~keV
band (2-eV display bins) with the best fit power-law model. No line is present at the
position of the Cd\,$L\alpha_1$ fluorescence line ($3.134$~keV).
The broad ($\sigma \simeq 25$~eV) excess around $3.12$~keV is thermal
S and Ar ejecta emission of the supernova remnants SNR~0509$-$68.7 and
G292.0$+$1.8, located in three fields of the sample. Lower panel:
residuals in units of $1\sigma$.}
\label{fig:cd_la1}
\end{figure}
The joint fit of Sec.~\ref{sm:fits} is the rigorous version of
the same idea, with exact rather than integer shifts and with the
continuum of every observation fitted separately. Adding the complex
to it as a second common component, an unresolved line at
$3.5089$~keV in every observation with one free non-negative flux,
gives $\Delta C = 2.5$--$2.9$ for the residual with the $S_{\dm}$-tied fluxes ($2$~eV and unresolved line) and $2.0$--$2.1$ for a common flux without the column density weights, and $\Delta C = 5.0$ for the complex (last row of Table~\ref{tab:residual}). The joint fit thus demonstrates the residual at $1.5$--$1.7\sigma$, against $2.5$--$3.0\sigma$ in the co-added spectrum, while both methods agree on the complex at $2.2\sigma$. Section~\ref{sm:origin} traces this difference to the
observations that carry the residual.

  \begin{table}[t!]
  \centering
    \caption{Trial factor for the residual of the Doppler corrected
  co-added spectrum ($\Delta C = 8.81$, local significance $3.0\sigma$, one sided $p = 1.5\times10^{-3}$). Each window is divided into independent line
  positions by the $5$~eV FWHM of the line spread function.}
  \setlength{\tabcolsep}{6pt}
  \begin{tabular}{lccc}
  \toprule
  Window & Width (eV) & Elements $N$ & Post trial \\
  \midrule
  Search band $3.48$--$3.56$~keV & 80 & 16 & $2.0\sigma$ \\
  $3.52 \pm 0.02$~keV, $1\sigma$ interval~\cite{Boyarsky:2014jta} & 40 & 8 & $2.3\sigma$ \\
  $3.539 \pm 0.011$~keV, $1\sigma$ interval~\cite{2015PhRvL.115p1301B} & 22 & 4.4 & $2.5\sigma$ \\
  \bottomrule
  \end{tabular}
    \label{tab:trials}  
  \end{table}
\smsubsection{Global significance of the residual}{sm:global}

The significances quoted so far are local. 
To evaluate the trial factors and thus the global significance, 
we consider two intervals of energies, both fixed before the 
fit: the $3.48$--$3.56$~keV search band, for a blind search, and the
energies compatible with the earlier detections of the $3.5$~keV line,
for a targeted test. In either region we count the independent line
positions as the width divided by the $5$~eV FWHM of the line spread
function (ten PI channels), the width of the unresolved line profile.  
The resulting global significance values are $2.0 - 2.5\sigma$ as shown in Table~\ref{tab:trials}.

\smsubsection{Instrumental and atomic counterparts}{sm:counterparts}

Two fluorescence lines of cadmium, Cd\,$L\beta_2$ and Cd\,$L\beta_{15}$,
both at $3.528$~keV~\citep{KortrightThompson:XDB,1967RvMP...39...78B}, lie
close to the $3.531$~keV residual. 
While the presence of cadmium in Resolve's NXB has not been reported,
whether these lines could produce the residual can be tested through the
companion Cd\,$L\alpha_1$ line at $3.134$~keV, expected to be $6.7$
times stronger than Cd\,$L\beta_{2,15}$~\citep{KortrightThompson:XDB}. 
The stacked spectrum in the $3.036$--$3.176$~keV interval around
Cd\,$L\alpha_1$, fitted with the power-law baseline model
($\Gamma_{\rm ph} = 1.46$ frozen), is shown in Fig.~\ref{fig:cd_la1}:
in contrast to the expected flux
($\simeq 8\times10^{-7}$~ph\,cm$^{-2}$\,s$^{-1}$, i.e.\ $6.7$ times
the flux of the $3.531$~keV excess), the fitted flux at
Cd\,$L\alpha_1$ is consistent with zero, with a $2\sigma$ upper limit~
of~$3.2\times10^{-7}$~ph\,cm$^{-2}$\,s$^{-1}$. The absence of
Cd\,$L\alpha_1$ rules out an instrumental origin of the reported
excess.

\smsection{The excess photons on the sky}{sm:origin}

In this section we discuss the distribution of the excess (above the background) photons at 3.5311~keV over the sky.
In order to identify the spatial origin of the photons, we merged the dataset of 60 observations into 28 fields, assigning to each field close in space observations. The spectrum of a field is defined as a stacked spectrum of all observations assigned to this field. We fit the spectra of the fields with the model of Sec.~\ref{sm:fits} with independent line amplitudes, each with the width $\sigma_E = 2$~eV at Doppler shifted positions
$3.5311$~keV ${}+ \Delta E_i$. 
The amplitudes are allowed to be negative~\cite{Protassov:2002sz}, with the only constraint that the model stays positive in every channel (which bounds the
amplitude from below in the fields with almost no counts).

\begin{figure}[t]
\centering
\includegraphics[width=\linewidth]{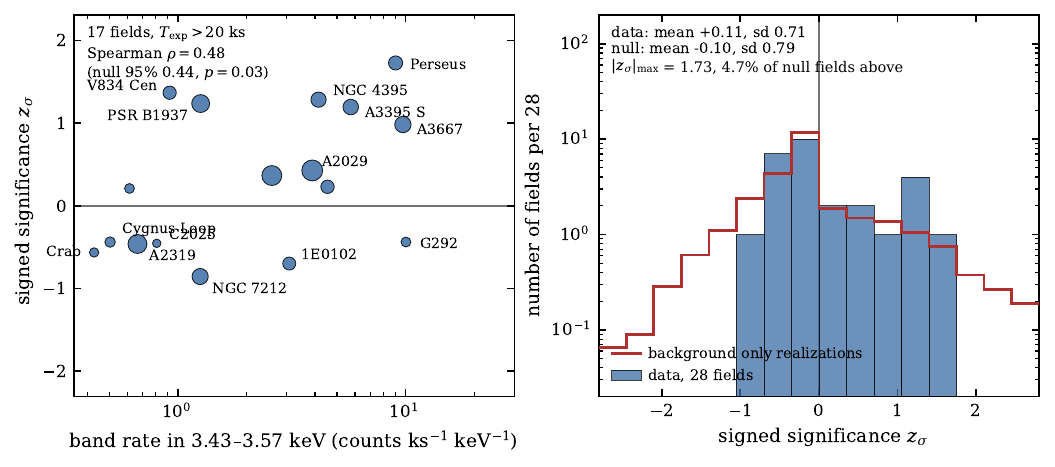}
\caption{\emph{Left:} signed significance $z_\sigma$ of the free line
amplitude of each sky field  against the field's
count rate in the fitting band, for the 17 fields with per-field exposure higher than 20~ks. Marker area is proportional to the exposure.
\emph{Right:} distribution of $z_\sigma$ over the 28 fields (bars) compared
with the distribution in background only realizations of the same fit
(step line, normalized to 28 fields); the negative tail of the
realizations comes from fields with few counts, whose amplitude runs
to the positivity bound of the model.}
\label{fig:field_z}
\end{figure}

\begin{figure}[t]
\centering
\includegraphics[width=\linewidth]{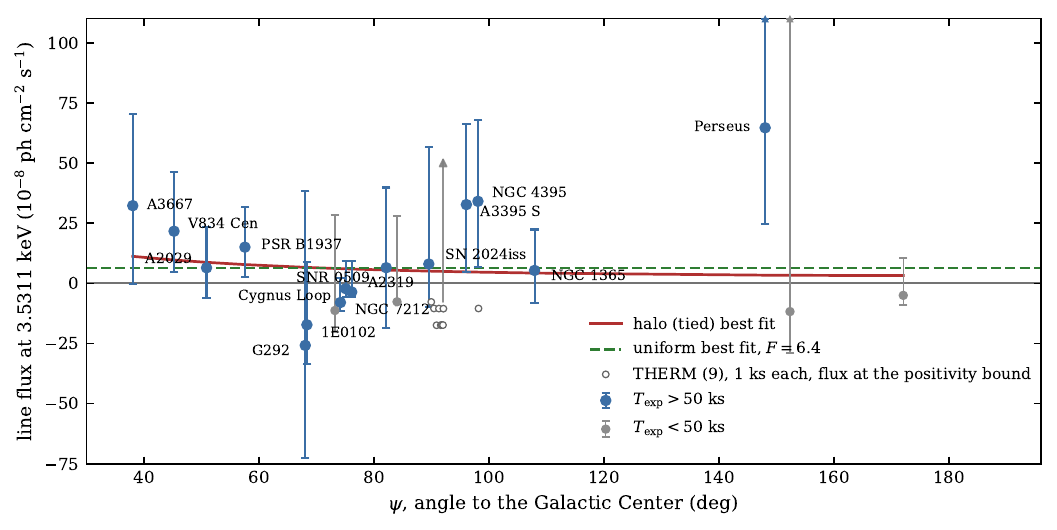}
\caption{Signed free line flux at $3.5311$~keV per sky field as a
function of the angle $\psi$ to the Galactic Center. Blue
points are fields with more than 50~ks of exposure and gray points
the others; error bars are the $\Delta C = 1$ intervals, and an
arrow marks an upper error bar extends above the panel. The nine
calibration pointings (open circles, 1~ks each) hold no counts and
sit at the positivity bound of the model. The red curve is the flux
of the tied fit at its best fit (joint fit row of
Table~\ref{tab:residual}), and the green dashed line the best fit uniform flux,
$6.4\times10^{-8}$~ph\,cm$^{-2}$\,s$^{-1}$.}
\label{fig:field_psi}
\end{figure}

\begin{figure}[t]
\centering
\includegraphics[width=\linewidth]{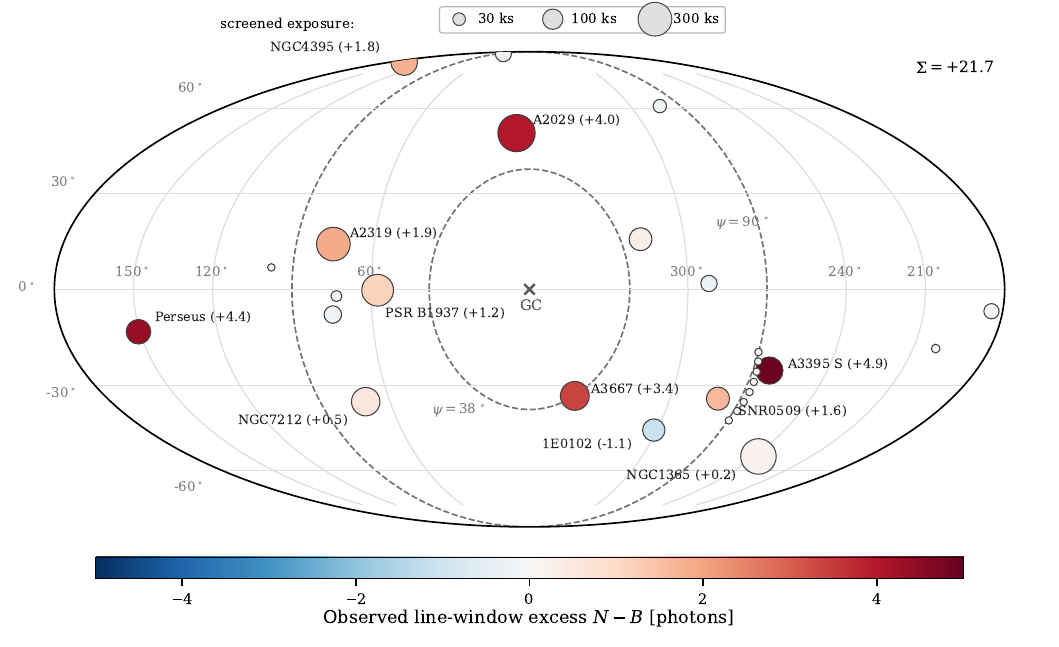}\\[2mm]
\includegraphics[width=\linewidth]{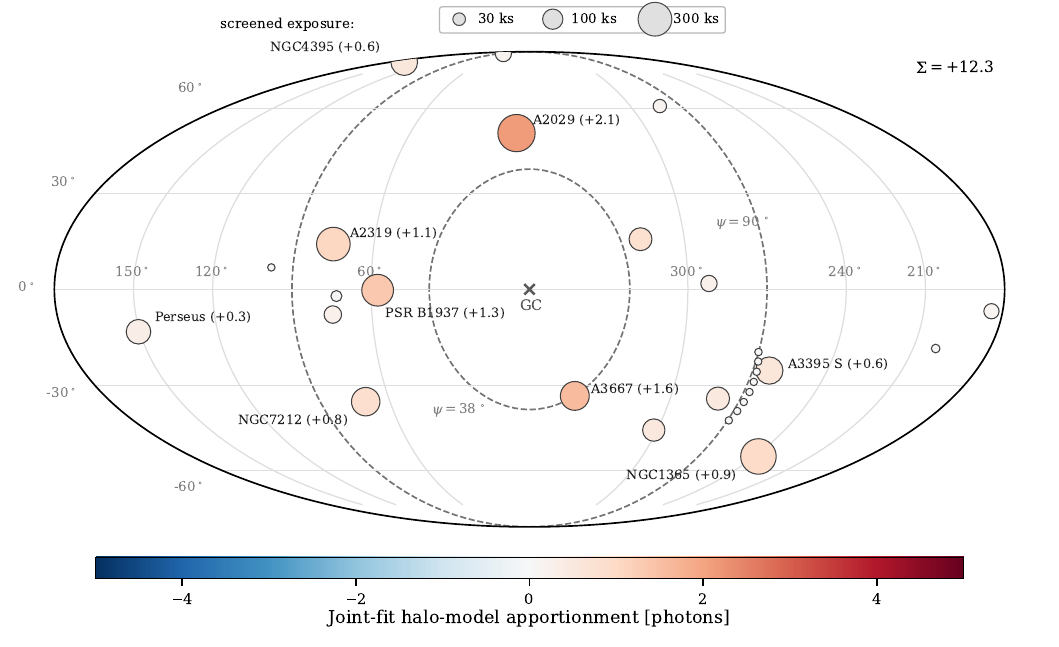}
\caption{Sky distribution (Galactic coordinates, Mollweide
projection) of the photon budget of the $3.531$~keV residual.
Marker area is proportional to the screened exposure; co-located
pointings are merged into single fields. The color scale is common
to both panels. \emph{Top:} observed excess counts $N-B$ in the
$\pm2.75$~eV (11 native 0.5~eV channels) Doppler shifted line
window of each field, relative to
the field's own fitted continuum ($\Sigma = +21.7$ photons; blue
fields show deficits). \emph{Bottom:} line photons apportioned by
the $S_{\dm}$-tied joint fit ($\Sigma = 12.3$ photons;
non-negative by construction). Dashed circles mark $\psi = 38^\circ$
(the sample boundary) and $\psi = 90^\circ$ from the
Galactic Center (GC); labels give per field photon totals for the
largest contributors and all deficits.}
\label{fig:sky_budget}
\end{figure}

Table~\ref{tab:fields} and Figs.~\ref{fig:field_psi}
and~\ref{fig:sky_budget} summarize the statistics of excess photons and their distribution over the sky. While in none of the fields the 3.5311~keV was detected with the significance exceeding $1.8\sigma$ (or below $-0.9\sigma$), in five fields (Perseus, ABELL3667, NGC4395, ABELL3395\_SOUTH and PSR\_B1937+21) the number of excess photons exceeds 2, see Tab.~\ref{tab:fields}. In seven fields we report the slight deficit of the expected number of photons.

The fields with the observed positive number of excess photons do not correlate with high DM column density in the Milky Way (see Fig.~\ref{fig:field_psi}) 
and are not located in any particular sky direction. We note however two potential correlations of the positive number of excess photons: \textit{(i)} with the high count rate fields; \textit{(ii)} the spatial correlation with clusters of galaxies. E.g. Perseus, ABELL3667, ABELL3395\_SOUTH and NGC4395 fields, demonstrating the positive number of excess photons are among the six fields with the highest count rate in the considered sample. Similarly, four positive number of photons excesses are observed in the direction of clusters of galaxies.

To assess the significance of the correlation between the positive excess and the count rate we performed 200 realizations of our dataset of fields around the background only model. For each realization of the dataset we performed a search of $\sigma_E=2$~eV width Gaussian line at the position of 3.5311~keV residual. For every field we defined the signed significance $z_\sigma$ 
of the detection of the line as the square root of the $\Delta C$. We define $z_\sigma$ to be positive for the lines with positive best-fit amplitudes and negative for the lines with negative best-fit amplitude. Fig.~\ref{fig:field_z} shows $z_\sigma$ for the real dataset. The left panel shows $z_\sigma$ as function of the fields count rate in the $3.43$--$3.57$~keV band, while the blue histogram in the right panel presents the distribution of signed significances. The red line in the right panel shows the similar distribution of the signed significances derived from the described above simulated datasets.

The real set of fields demonstrates the positive correlation with the $3.43$--$3.57$~keV band rate with the Spearman rank correlation of $r_s = 0.48$. In the background only realizations the distribution of the same quantity has a median of $0.11$ and a 95\% quantile of $0.44$. The p-value for $r_s = 0.48$ of the real data is $p = 0.025$ which corresponds to a marginal $\sim 2.2\sigma$ correlation significance. We note though that the highest count rate field (G292\_NW) shows a deficit ($-1.0$ photon) in the excess number of photons.

The possibility of a correlation between the positive excess and galaxy clusters can be also connected to the additional (to the MW) dark matter column density in clusters of galaxies and/or to the presence of hot cluster plasma astrophysical lines redshifted to the analyzed energy band. Table~\ref{tab:clusters} reports on the properties of the galaxy clusters in the fields with the positive count excess. It summarizes the name of the field, redshift of the corresponding cluster, offset of the observation with respect to the clusters center and ratio of the expected MW/cluster DM column densities for the corresponding field. The rest of the columns correspond to the \textit{observed }energy of a putative DM decay line located at $3.5311$~keV in the frame of a cluster and an energy at which an astrophysical ICM line shall be present to appear at $3.5311$~keV for the observer.  

This Table demonstrates that if the DM line is present in clusters at $3.5311$~keV, due to a clusters redshift it is shifted well below the analyzed energy range. The astrophysical ICM lines can contribute to the excess of photons only if they are located in $3.594-3.804$~keV energy band.

At similar energies the AtomDB list~\cite{Foster:2012hy} contains the \ion{Cl}{XVII} Lyman series (Ly$\delta$ at $3.571$~keV, Ly$\epsilon$ at $3.616$~keV, series limit at $3.720$~keV), \ion{Ar}{XVII} He$\beta$ ($3.686$~keV), the \ion{K}{XIX} Ly$\alpha$ doublet ($3.699$ and $3.706$~keV),
\ion{Ca}{XIX} He$\alpha$ ($3.861$--$3.902$~keV) and \ion{Ar}{XVIII} Ly$\beta$ ($3.936$~keV). The rest energies of Table~\ref{tab:clusters} fall in the gap between \ion{K}{XIX}
Ly$\alpha$ and \ion{Ca}{XIX} He$\alpha$, where the only feature is
the \ion{Cl}{XVII} series limit, a recombination edge broadened to
the gas temperature rather than a line. Redshifted into the band,
the nearest ICM lines would appear at $3.5196$ and $3.5134$~keV (\ion{K}{XIX} Ly$\alpha$ from A3395S, $11.5$ and $17.7$~eV below the residual),
at $3.5106$ and $3.5045$~keV (the same doublet from A3667), at
$3.5005$ and $3.4916$~keV (\ion{Ar}{XVII} He$\beta$ from A3395S and
A3667), and at $3.508$ and $3.553$~keV (\ion{Cl}{XVII} Ly$\delta$
and Ly$\epsilon$ from Perseus); the \ion{K}{XIX} lines of A2029
land at $3.440$~keV. None of these lines is within two resolution elements
of $3.5311$~keV. 

We conclude thus on the unlikely possibility of the correlation of the positive number of photons excess with clusters of galaxies due to the increased DM column density or presence of ICM astrophysical lines originated from these objects.

Figure~\ref{fig:sky_budget} shows the sky distribution of the
photon budget of the $3.531$~keV residual assuming the free line normalisations (top panel) and line normalizations tied to the DM column density (bottom panel).  

For each field the top panel gives
the observed excess $N-B$ in the $\pm2.75$~eV window around the
Doppler shifted $3.5311$~keV line energy of the field, where $N$ is
the observed count and $B$ the prediction of the field's own fitted
continuum; the panels share a common color scale. The observed
excess ($\Sigma = +21.7$ photons) is spread mainly over four fields:
ABELL3395\_SOUTH ($\psi = 96^\circ$, $+4.9$ photons), Perseus
($\psi = 148^\circ$, $+4.4$), Abell2029 ($\psi = 51^\circ$, $+4.0$)
and ABELL3667 ($\psi = 38^\circ$, $+3.4$), the first two of low and
the last two of high column density, while the only deficit is
1E0102 ($-1.1$). 

The bottom panel shows, for
comparison, the line photons apportioned by the $S_{\dm}$-tied joint
fit ($\Sigma = 12.3$; non-negative by construction).
The spatial
distribution of the excess counts shows no clear preference for
sight lines of high column density that a Milky Way halo DM interpretation
would predict, see also Fig.~\ref{fig:field_psi}.

\begin{table*}[t]
\centering
\caption{Signed free line flux at $3.5311$~keV per sky field, for the
17 fields with more than 20~ks of exposure, ordered by the angle
$\psi$ to the Galactic Center. Pointings of the same target within
one degree are merged. Columns: number of pointings, Galactic
coordinates, $\psi$, column density $S_{\dm}$ in
$10^{22}~\mathrm{GeV\,cm^{-2}}$, screened exposure, counts in the
fitting band, the fitted flux with $\Delta C = 1$ errors
in $10^{-8}$~ph\,cm$^{-2}$\,s$^{-1}$ (an error marked with an
asterisk is cut by the positivity bound of the model rather than by
$\Delta C = 1$), the line photons at the best fit, and the flux that
the tied fit of Sec.~\ref{sm:fits} assigns to the field at its
own best fit (joint fit row of Table~\ref{tab:residual}), that is,
the expectation if the whole
excess is a halo line. The 11 fields with less than 20~ks (the nine
calibration pointings, Cygnus X-1 and $\psi$~Orionis) are found to be close to the
positivity bound with $|z_\sigma| < 0.6$.}
\label{tab:fields}
\begin{tabular}{lrrrrrrrrrr}
\toprule
field & $n$ & $l$ & $b$ & $\psi$ & $S_{\dm}$ & $T_{\rm exp}$ (ks) & counts & flux & photons & tied \\
\midrule
ABELL3667\_CF\_IN & 1 & 340.8 & $-33.5$ & 38.0 & 3.67 & 216 & 296 & $+32^{+38}_{-33}$ & $+4.6$ & 11.2 \\
V834\_Cen & 1 & 317.0 & 15.5 & 45.2 & 3.17 & 131 & 17 & $+22^{+25}_{-17}$ & $+1.8$ & 9.7 \\
Abell2029 & 5 & 6.5 & 50.5 & 50.9 & 2.86 & 370 & 203 & $+6^{+17}_{-13}$ & $+1.6$ & 8.7 \\
PSR\_B1937+21 & 2 & 57.5 & $-0.3$ & 57.5 & 2.55 & 265 & 47 & $+15^{+17}_{-13}$ & $+2.6$ & 7.8 \\
G292\_NW & 2 & 292.0 & 1.8 & 68.0 & 2.17 & 59 & 84 & $-26^{+64}_{-47}$ & $-1.0$ & 6.6 \\
1E0102.2$-$7219 & 2 & 301.6 & $-45.1$ & 68.3 & 2.17 & 124 & 54 & $-17^{+26}_{-17}$ & $-1.4$ & 6.6 \\
NGC7212 & 5 & 70.4 & $-35.4$ & 74.1 & 2.00 & 209 & 37 & $-8^{+10}_{-3}$ & $-1.1$ & 6.1 \\
Cygnus\_Loop & 2 & 74.9 & $-7.8$ & 75.1 & 1.98 & 70 & 5 & $-2^{+12}_{-4}$ & $-0.1$ & 6.0 \\
Abell2319\_BS & 6 & 75.7 & 13.9 & 76.1 & 1.95 & 299 & 28 & $-4^{+13}_{-2}$ & $-0.7$ & 6.0 \\
SNR0509$-$68.7 & 1 & 279.6 & $-34.4$ & 82.1 & 1.81 & 134 & 86 & $+6^{+33}_{-25}$ & $+0.6$ & 5.5 \\
C2025\_N1 & 11 & 282.4 & 60.8 & 84.0 & 1.77 & 35 & 4 & $-8^{+36}_{-0^{*}}$ & $-0.2$ & 5.4 \\
SN2024iss & 1 & 78.2 & 87.6 & 89.5 & 1.67 & 58 & 5 & $+8^{+49}_{-18}$ & $+0.3$ & 5.1 \\
ABELL3395\_SOUTH & 1 & 263.3 & $-25.3$ & 96.0 & 1.56 & 191 & 155 & $+33^{+33}_{-28}$ & $+4.1$ & 4.8 \\
NGC4395 & 1 & 162.1 & 81.5 & 98.1 & 1.53 & 183 & 107 & $+34^{+34}_{-27}$ & $+4.1$ & 4.7 \\
NGC1365 & 2 & 238.0 & $-54.6$ & 107.9 & 1.40 & 335 & 122 & $+5^{+17}_{-14}$ & $+1.2$ & 4.3 \\
Perseus (CCD3, CCD4) & 4 & 150.5 & $-13.1$ & 148.0 & 1.11 & 152 & 194 & $+65^{+48}_{-40}$ & $+6.5$ & 3.4 \\
Crab & 2 & 184.3 & $-6.7$ & 172.0 & 1.05 & 50 & 3 & $-5^{+16}_{-4^{*}}$ & $-0.2$ & 3.2 \\
\bottomrule
\end{tabular}
\end{table*}

\begin{table*}[t]
\centering
\caption{The clusters in the four fields with the positive number of excess photons. Columns:
the field of Fig.~\ref{fig:sky_budget}; the cluster redshift $z$
(the ICM value of Ref.~\cite{Ota:2026nsr} for A3395S, the NED
values otherwise); the projected distance of the pointing from the
cluster center; the DM column density of the cluster along the
pointing, $S_{\rm cl}$, in units of the Milky Way column density
$S_{\dm}$ of the same pointing (Table~\ref{tab:dm_params}), for an NFW halo of
mass $M_{200}$ and concentration $c$ (Perseus: $6.7\times10^{14}$
$M_\odot$, $c = 5$~\cite{Simionescu:2011ii}; A2029 and A3667:
$10^{15}$ $M_\odot$; A3395S: $3\times10^{14}$ $M_\odot$;
$c = 4$ otherwise; masses rounded from Ref.~\cite{Piffaretti:2010my},
uncertain by a factor of about two), integrated to $r_{200}$ and
averaged over the $3'\times3'$ field of view; the
detector frame energy at which a DM line of rest energy
$3.5311$~keV from the cluster would appear, $3.5311/(1+z)$~keV; the
rest energy an ICM line must have to appear at $3.5311$~keV,
$3.5311\,(1+z)$~keV.
Where a field merges several pointings the range over the pointings
is given.}
\label{tab:clusters}
\setlength{\tabcolsep}{5pt}
\begin{tabular}{lccccc}
\toprule
field & $z$ & offset & $S_{\rm cl}/S_{\dm}$ & cluster DM line at & ICM line needs  \\
 & & & & [keV] & [keV]   \\
\midrule
ABELL3395\_SOUTH & 0.0529 & core & 5--7 & 3.354 & 3.718  \\
Perseus & 0.0179 & $11'$--$18'$ & 2.5--4.5 & 3.469 & 3.594  \\
Abell2029 & 0.0773 & $6'$--$11'$ & 0.4--0.9 & 3.278 & 3.804  \\
ABELL3667 & 0.0556 & $5'$ & 1.3 & 3.345 & 3.727  \\
\bottomrule
\end{tabular}
\end{table*}

\begin{table*}[tp]
\centering
\scriptsize
\setlength{\tabcolsep}{2pt}
\caption{XRISM/Resolve sample ($R < 0.136$~cts\,s$^{-1}$,
$T > 1000$\,s, $n = 60$):
$\sum T = 3{,}797{,}911.7$\,s (ONTIME),
$\sum T_{\rm exp} = 2{,}916{,}957.1$\,s (fitted exposure). Columns: ObsID,
target name, equatorial coordinates (J2000), Galactic coordinates, angle to
the Galactic Center $\psi$, cleaned event ONTIME $T$, screened exposure
$T_{\rm exp}$ (EXPOSURE keyword of the spectrum; CORTIME$>8$ GTI), and total
count rate $R$ of the archival cleaned Resolve
event list. This is the sample of Sec.~\ref{sm:dataset}.}
\label{tab:obs}
\begin{tabular*}{\textwidth}{rlrrrrrrrrr}
\toprule
\# & ObsID & Name~~~~~~~~~~~ & RA (deg) & Dec (deg) & $l$ (deg) & $b$ (deg) & $\psi$ (deg) & $T$ (s) & $T_{\rm exp}$ (s) & $R$ (cts\,s$^{-1}$) \\
\midrule
1 & 001004010 & RESOLVE\_THERM3\_S & 97.594 & -58.979 & 268.232 & -25.649 & 91.594 & 1249.8 & 1249.8 & 0.0464 \\
2 & 001003010 & RESOLVE\_THERM2\_S & 104.106 & -58.983 & 269.073 & -22.384 & 90.857 & 1249.8 & 1249.8 & 0.0496 \\
3 & 001002010 & RESOLVE\_THERM1\_S & 110.165 & -58.936 & 270.07 & -19.402 & 89.934 & 1250.2 & 1250.2 & 0.0512 \\
4 & 001007010 & RESOLVE\_THERM6\_S & 78.418 & -59.027 & 267.805 & -35.484 & 91.787 & 1250.0 & 1011.4 & 0.0576 \\
5 & 902001110 & C2025\_N1 & 182.739 & -0.351 & 281.788 & 60.848 & 84.289 & 3556.0 & 3556.0 & 0.0585 \\
6 & 001006010 & RESOLVE\_THERM5\_S & 84.817 & -59.008 & 267.583 & -32.197 & 92.045 & 1250.6 & 1030.9 & 0.0648 \\
7 & 902001100 & C2025\_N1 & 182.854 & -0.392 & 282.038 & 60.846 & 84.169 & 3543.7 & 3132.7 & 0.0652 \\
8 & 001009010 & RESOLVE\_THERM8\_S & 66.052 & -59.052 & 269.462 & -41.704 & 90.401 & 1249.8 & 1007.4 & 0.0664 \\
9 & 000153300 & Abell2319\_BS & 289.762 & 44.105 & 75.673 & 13.919 & 76.102 & 26861.1 & 26158.5 & 0.0675 \\
10 & 001008010 & RESOLVE\_THERM7\_S & 72.459 & -59.031 & 268.375 & -38.516 & 91.271 & 1250.1 & 998.4 & 0.0776 \\
11 & 902001040 & C2025\_N1 & 183.527 & -0.63 & 283.503 & 60.826 & 83.464 & 2764.4 & 2764.4 & 0.0807 \\
12 & 000122000 & PSR\_B1937+21\_OFFSET & 294.914 & 21.622 & 57.544 & -0.273 & 57.545 & 58114.5 & 51259.5 & 0.0812 \\
13 & 902001120 & C2025\_N1 & 182.64 & -0.316 & 281.573 & 60.849 & 84.392 & 4169.0 & 4019.0 & 0.0813 \\
14 & 000135000 & RESOLVE\_THERM5 & 319.328 & 58.906 & 98.21 & 6.735 & 98.153 & 1040.1 & 1040.1 & 0.0827 \\
15 & 902001050 & C2025\_N1 & 183.418 & -0.591 & 283.265 & 60.83 & 83.579 & 4993.7 & 4204.7 & 0.0841 \\
16 & 001005010 & RESOLVE\_THERM4\_S & 91.201 & -58.997 & 267.742 & -28.913 & 91.976 & 1240.9 & 1047.3 & 0.0846 \\
17 & 000123000 & PSR\_B1937+21 & 294.91 & 21.583 & 57.508 & -0.289 & 57.509 & 245115.4 & 213289.4 & 0.0874 \\
18 & 201110040 & NGC7212 & 331.756 & 10.231 & 70.419 & -35.378 & 74.142 & 62005.2 & 55392.2 & 0.0876 \\
19 & 001010010 & PSI\_ORIONIS\_CLUSTER & 81.249 & 1.7 & 201.124 & -18.307 & 152.324 & 9079.2 & 6348.3 & 0.0888 \\
20 & 902001130 & C2025\_N1 & 182.528 & -0.277 & 281.328 & 60.85 & 84.509 & 5937.2 & 4153.9 & 0.0889 \\
21 & 100007010 & Crab\_stray\_strong & 82.591 & 21.752 & 184.256 & -6.741 & 172.033 & 8863.1 & 5632.1 & 0.0898 \\
22 & 902001060 & C2025\_N1 & 183.307 & -0.552 & 283.024 & 60.834 & 83.695 & 5891.7 & 3859.0 & 0.0903 \\
23 & 000148000 & Abell2029\_SW & 227.575 & 5.653 & 6.209 & 50.617 & 50.892 & 27591.0 & 18850.5 & 0.0908 \\
24 & 000114000 & Abell2319\_BS2 & 289.69 & 44.164 & 75.707 & 13.99 & 76.139 & 145436.0 & 108358.0 & 0.0922 \\
25 & 100008010 & Cygnus\_Loop & 312.705 & 31.89 & 74.921 & -7.768 & 75.062 & 45870.0 & 34368.5 & 0.0935 \\
26 & 201110050 & NGC7212 & 331.756 & 10.231 & 70.419 & -35.378 & 74.142 & 32699.2 & 25715.2 & 0.0935 \\
27 & 201110020 & NGC7212 & 331.756 & 10.231 & 70.419 & -35.378 & 74.142 & 43805.0 & 37818.0 & 0.0936 \\
28 & 900002010 & SN2024iss & 194.775 & 28.811 & 78.222 & 87.614 & 89.513 & 68219.6 & 57798.6 & 0.0939 \\
29 & 100008020 & Cygnus\_Loop & 312.705 & 31.89 & 74.921 & -7.768 & 75.062 & 49058.0 & 35930.5 & 0.0946 \\
30 & 201110030 & NGC7212 & 331.756 & 10.231 & 70.419 & -35.378 & 74.142 & 41742.4 & 36978.4 & 0.0954 \\
31 & 100007020 & Crab\_stray\_weak & 82.591 & 21.752 & 184.256 & -6.741 & 172.033 & 61791.8 & 43953.1 & 0.0963 \\
32 & 000159000 & V834\_Cen & 212.28 & -45.288 & 316.978 & 15.455 & 45.198 & 180481.9 & 130972.4 & 0.1014 \\
33 & 000153100 & Abell2319\_BS & 289.762 & 44.105 & 75.673 & 13.919 & 76.102 & 2778.0 & 2778.0 & 0.1022 \\
34 & 101018010 & Perseus\_CCD3\_SPRING & 50.264 & 41.703 & 150.668 & -12.973 & 148.162 & 43006.9 & 36529.9 & 0.1034 \\
35 & 000146000 & Abell2029\_N & 227.765 & 5.858 & 6.646 & 50.588 & 50.904 & 19061.0 & 13685.0 & 0.1043 \\
36 & 000108000 & Cygnus\_X-1\_BS & 306.041 & 33.911 & 73.167 & -2.084 & 73.178 & 32431.0 & 19787.0 & 0.1047 \\
37 & 102012010 & Perseus\_CCD3\_SPRING & 50.265 & 41.703 & 150.668 & -12.972 & 148.163 & 39734.7 & 33102.7 & 0.1049 \\
38 & 201110010 & NGC7212 & 331.756 & 10.231 & 70.419 & -35.378 & 74.142 & 71074.7 & 53465.7 & 0.1055 \\
39 & 300053010 & Abell2029\_N2 & 227.794 & 5.825 & 6.631 & 50.545 & 50.86 & 366150.1 & 287792.2 & 0.1058 \\
40 & 902001140 & C2025\_N1 & 182.416 & -0.237 & 281.084 & 60.85 & 84.627 & 6373.5 & 3086.5 & 0.1062 \\
41 & 902001070 & C2025\_N1 & 183.197 & -0.513 & 282.784 & 60.838 & 83.81 & 5950.2 & 3079.0 & 0.1064 \\
42 & 000147000 & Abell2029\_E & 227.848 & 5.714 & 6.542 & 50.437 & 50.744 & 23025.6 & 17707.1 & 0.1082 \\
43 & 000118000 & 1E0102.2-7219 & 16.005 & -72.033 & 301.559 & -45.062 & 68.304 & 94469.9 & 70698.9 & 0.1119 \\
44 & 000136000 & 1E0102.2-7219 & 16.01 & -72.031 & 301.556 & -45.063 & 68.306 & 75356.2 & 53434.7 & 0.112 \\
45 & 902001080 & C2025\_N1 & 183.086 & -0.474 & 282.543 & 60.841 & 83.926 & 4764.0 & 2102.4 & 0.1138 \\
46 & 000100000 & Abell2319\_BS & 289.755 & 44.083 & 75.651 & 13.914 & 76.081 & 9105.4 & 6854.4 & 0.1148 \\
47 & 201020010 & ABELL3395\_SOUTH & 96.702 & -54.546 & 263.312 & -25.266 & 96.046 & 266408.4 & 191382.5 & 0.1197 \\
48 & 000152000 & Abell2029\_N2 & 227.794 & 5.825 & 6.631 & 50.546 & 50.86 & 48124.6 & 31912.2 & 0.1205 \\
49 & 201048010 & NGC4395 & 186.455 & 33.547 & 162.088 & 81.533 & 98.054 & 231158.0 & 182951.9 & 0.121 \\
50 & 300075010 & NGC1365 & 53.403 & -36.14 & 237.955 & -54.597 & 107.901 & 200540.7 & 146433.8 & 0.1233 \\
51 & 102013010 & Perseus\_CCD4\_SPRING & 49.831 & 41.676 & 150.406 & -13.174 & 147.851 & 45718.5 & 39137.6 & 0.1260 \\
52 & 201050010 & ABELL3667\_CF\_IN & 303.265 & -56.886 & 340.795 & -33.466 & 38.021 & 275635.3 & 215807.2 & 0.1264 \\
53 & 300075020 & NGC1365 & 53.401 & -36.141 & 237.957 & -54.599 & 107.9 & 259186.5 & 188071.0 & 0.1266 \\
54 & 101019010 & Perseus\_CCD4\_SPRING & 49.824 & 41.675 & 150.402 & -13.177 & 147.846 & 50195.0 & 43502.5 & 0.1278 \\
55 & 201104010 & SNR0509-68.7 & 77.249 & -68.727 & 279.581 & -34.367 & 82.103 & 180756.7 & 134305.7 & 0.1278 \\
56 & 201096010 & G292\_NW & 171.091 & -59.225 & 291.988 & 1.784 & 68.023 & 40599.9 & 29663.5 & 0.1303 \\
57 & 000104000 & Abell2319\_BS & 289.76 & 44.104 & 75.672 & 13.92 & 76.101 & 101904.5 & 71400.0 & 0.1313 \\
58 & 201096020 & G292\_NW & 171.091 & -59.225 & 291.988 & 1.784 & 68.023 & 40152.4 & 29716.4 & 0.1326 \\
59 & 902001030 & C2025\_N1 & 183.621 & -0.663 & 283.707 & 60.822 & 83.366 & 1158.9 & 1144.4 & 0.1338 \\
60 & 000104200 & Abell2319\_BS & 289.76 & 44.104 & 75.672 & 13.92 & 76.101 & 114470.7 & 83026.7 & 0.1352 \\
\bottomrule
\end{tabular*}
\end{table*}

\begin{table*}[tp]
\centering
\scriptsize
\setlength{\tabcolsep}{2pt}
\caption{Dark matter signal parameters for the XRISM/Resolve sample,
sorted by count rate $R$ (ascending). Computed for $m_{\dm} = 7$\,keV and
$\sin^2(2\theta) = 2.0\times10^{-10}$ assuming an NFW halo profile. 
$v_{\rm los}$ and $E_{\rm obs}$ include the solar motion and the heliocentric
orbital velocity of the Earth at the epoch of each observation ($v_\oplus$);
$N_\gamma$ is computed with the screened spectrum exposure $T_{\rm exp}$
(CORTIME$>8$ GTI) and $A_{\rm eff}=65$\,cm$^2$. For the 60 observation
sample $\sum N_\gamma = 18.073$. Sky positions and geometry are in Table~\ref{tab:obs}; see Sec.~\ref{sm:dataset}.}
\label{tab:dm_params}
\begin{tabular*}{\textwidth}{@{\extracolsep{\fill}}rllccccccc}
\toprule
\# & ObsID & Name & $v_{\rm los}$ & $v_\oplus$ & $E_{\rm obs}$ & $\Delta E$ & $S_{\dm}$ & $F_\gamma$ & $N_\gamma$ \\
 & & & (km\,s$^{-1}$) & (km\,s$^{-1}$) & (keV) & (eV) & ($10^{22}$\,GeV\,cm$^{-2}$) & ($10^{-8}$\,ph\,cm$^{-2}$\,s$^{-1}$) & \\
\midrule
1 & 001004010 & RESOLVE\_THERM3\_S & $-$217.50 & $+$3.3 & 3.497461 & $-$2.539 & 1.628 & 7.293 & 0.00592 \\
2 & 001003010 & RESOLVE\_THERM2\_S & $-$223.35 & $+$2.6 & 3.497393 & $-$2.607 & 1.641 & 7.350 & 0.00597 \\
3 & 001002010 & RESOLVE\_THERM1\_S & $-$228.15 & $+$1.8 & 3.497336 & $-$2.664 & 1.657 & 7.422 & 0.00603 \\
4 & 001007010 & RESOLVE\_THERM6\_S & $-$196.84 & $+$4.0 & 3.497702 & $-$2.298 & 1.625 & 7.278 & 0.00479 \\
5 & 902001110 & C2025\_N1 & $-$80.49 & $+$27.1 & 3.499060 & $-$0.940 & 1.764 & 7.902 & 0.01826 \\
6 & 001006010 & RESOLVE\_THERM5\_S & $-$204.21 & $+$4.0 & 3.497616 & $-$2.384 & 1.621 & 7.259 & 0.00487 \\
7 & 902001100 & C2025\_N1 & $-$80.43 & $+$27.1 & 3.499061 & $-$0.939 & 1.767 & 7.913 & 0.01612 \\
8 & 001009010 & RESOLVE\_THERM8\_S & $-$181.65 & $+$3.3 & 3.497879 & $-$2.121 & 1.649 & 7.385 & 0.00484 \\
9 & 000153300 & Abell2319\_BS & $+$229.73 & $-$1.6 & 3.502682 & $+$2.682 & 1.949 & 8.729 & 0.14844 \\
10 & 001008010 & RESOLVE\_THERM7\_S & $-$189.64 & $+$3.8 & 3.497786 & $-$2.214 & 1.634 & 7.318 & 0.00475 \\
11 & 902001040 & C2025\_N1 & $-$80.06 & $+$26.7 & 3.499065 & $-$0.935 & 1.781 & 7.978 & 0.01433 \\
12 & 000122000 & PSR\_B1937+21\_OFFSET & $+$190.28 & $-$19.2 & 3.502222 & $+$2.221 & 2.549 & 11.419 & 0.38043 \\
13 & 902001120 & C2025\_N1 & $-$80.54 & $+$27.2 & 3.499060 & $-$0.940 & 1.762 & 7.892 & 0.02062 \\
14 & 000135000 & RESOLVE\_THERM5 & $+$224.51 & $-$11.9 & 3.502621 & $+$2.621 & 1.525 & 6.831 & 0.00462 \\
15 & 902001050 & C2025\_N1 & $-$80.12 & $+$26.7 & 3.499065 & $-$0.935 & 1.779 & 7.967 & 0.02178 \\
16 & 001005010 & RESOLVE\_THERM4\_S & $-$211.12 & $+$3.8 & 3.497535 & $-$2.465 & 1.622 & 7.264 & 0.00495 \\
17 & 000123000 & PSR\_B1937+21 & $+$190.82 & $-$18.6 & 3.502228 & $+$2.228 & 2.551 & 11.426 & 1.58420 \\
18 & 201110040 & NGC7212 & $+$158.78 & $-$25.4 & 3.501854 & $+$1.854 & 1.999 & 8.954 & 0.32240 \\
19 & 001010010 & PSI\_ORIONIS\_CLUSTER & $-$67.26 & $+$27.4 & 3.499215 & $-$0.785 & 1.094 & 4.899 & 0.02022 \\
20 & 902001130 & C2025\_N1 & $-$80.60 & $+$27.2 & 3.499059 & $-$0.941 & 1.76 & 7.882 & 0.02129 \\
21 & 100007010 & Crab\_stray\_strong & $-$59.36 & $-$29.7 & 3.499307 & $-$0.693 & 1.053 & 4.717 & 0.01727 \\
22 & 902001060 & C2025\_N1 & $-$80.18 & $+$26.8 & 3.499064 & $-$0.936 & 1.776 & 7.956 & 0.01995 \\
23 & 000148000 & Abell2029\_SW & $+$54.59 & $+$25.4 & 3.500637 & $+$0.637 & 2.855 & 12.787 & 0.15670 \\
24 & 000114000 & Abell2319\_BS2 & $+$219.22 & $-$12.0 & 3.502559 & $+$2.559 & 1.948 & 8.725 & 0.61458 \\
25 & 100008010 & Cygnus\_Loop & $+$250.18 & $+$17.5 & 3.502921 & $+$2.921 & 1.975 & 8.847 & 0.19763 \\
26 & 201110050 & NGC7212 & $+$159.27 & $-$24.9 & 3.501859 & $+$1.859 & 1.999 & 8.954 & 0.14967 \\
27 & 201110020 & NGC7212 & $+$157.48 & $-$26.7 & 3.501839 & $+$1.839 & 1.999 & 8.954 & 0.22011 \\
28 & 900002010 & SN2024iss & $-$5.00 & $-$22.2 & 3.499942 & $-$0.058 & 1.665 & 7.455 & 0.28020 \\
29 & 100008020 & Cygnus\_Loop & $+$250.35 & $+$17.7 & 3.502923 & $+$2.923 & 1.975 & 8.847 & 0.20661 \\
30 & 201110030 & NGC7212 & $+$158.30 & $-$25.9 & 3.501848 & $+$1.848 & 1.999 & 8.954 & 0.21522 \\
31 & 100007020 & Crab\_stray\_weak & $-$59.30 & $-$29.7 & 3.499308 & $-$0.692 & 1.053 & 4.717 & 0.13476 \\
32 & 000159000 & V834\_Cen & $-$122.91 & $+$26.0 & 3.498565 & $-$1.435 & 3.172 & 14.208 & 1.20961 \\
33 & 000153100 & Abell2319\_BS & $+$228.87 & $-$2.4 & 3.502672 & $+$2.672 & 1.949 & 8.729 & 0.01576 \\
34 & 101018010 & Perseus\_CCD3\_SPRING & $+$77.54 & $-$26.6 & 3.500905 & $+$0.905 & 1.109 & 4.965 & 0.11795 \\
35 & 000146000 & Abell2029\_N & $+$55.46 & $+$25.1 & 3.500647 & $+$0.647 & 2.854 & 12.785 & 0.11372 \\
36 & 000108000 & Cygnus\_X-1\_BS & $+$215.11 & $-$18.6 & 3.502511 & $+$2.511 & 2.025 & 9.069 & 0.11666 \\
37 & 102012010 & Perseus\_CCD3\_SPRING & $+$76.81 & $-$27.3 & 3.500897 & $+$0.897 & 1.109 & 4.965 & 0.10689 \\
38 & 201110010 & NGC7212 & $+$156.53 & $-$27.6 & 3.501827 & $+$1.827 & 1.999 & 8.954 & 0.31119 \\
39 & 300053010 & Abell2029\_N2 & $+$3.32 & $-$27.0 & 3.500039 & $+$0.039 & 2.857 & 12.795 & 2.39398 \\
40 & 902001140 & C2025\_N1 & $-$80.66 & $+$27.3 & 3.499058 & $-$0.942 & 1.757 & 7.871 & 0.01579 \\
41 & 902001070 & C2025\_N1 & $-$80.24 & $+$26.9 & 3.499063 & $-$0.937 & 1.774 & 7.946 & 0.01590 \\
42 & 000147000 & Abell2029\_E & $+$55.35 & $+$25.2 & 3.500646 & $+$0.646 & 2.862 & 12.821 & 0.14755 \\
43 & 000118000 & 1E0102.2-7219 & $-$158.59 & $-$12.4 & 3.498149 & $-$1.851 & 2.165 & 9.697 & 0.44566 \\
44 & 000136000 & 1E0102.2-7219 & $-$154.61 & $-$8.4 & 3.498195 & $-$1.805 & 2.165 & 9.696 & 0.33683 \\
45 & 902001080 & C2025\_N1 & $-$80.30 & $+$26.9 & 3.499063 & $-$0.938 & 1.772 & 7.935 & 0.01085 \\
46 & 000100000 & Abell2319\_BS & $+$219.21 & $-$12.1 & 3.502559 & $+$2.559 & 1.949 & 8.731 & 0.03890 \\
47 & 201020010 & ABELL3395\_SOUTH & $-$223.99 & $-$3.0 & 3.497385 & $-$2.615 & 1.557 & 6.972 & 0.86760 \\
48 & 000152000 & Abell2029\_N2 & $+$56.50 & $+$26.2 & 3.500660 & $+$0.660 & 2.857 & 12.794 & 0.26546 \\
49 & 201048010 & NGC4395 & $+$37.80 & $+$21.3 & 3.500441 & $+$0.441 & 1.527 & 6.838 & 0.81341 \\
50 & 300075010 & NGC1365 & $-$144.08 & $-$16.3 & 3.498318 & $-$1.682 & 1.399 & 6.266 & 0.59647 \\
51 & 102013010 & Perseus\_CCD4\_SPRING & $+$77.79 & $-$27.2 & 3.500908 & $+$0.908 & 1.11 & 4.971 & 0.12646 \\
52 & 201050010 & ABELL3667\_CF\_IN & $-$84.35 & $-$22.9 & 3.499015 & $-$0.985 & 3.67 & 16.436 & 2.30602 \\
53 & 300075020 & NGC1365 & $-$111.47 & $+$16.3 & 3.498699 & $-$1.301 & 1.399 & 6.266 & 0.76607 \\
54 & 101019010 & Perseus\_CCD4\_SPRING & $+$78.22 & $-$26.7 & 3.500913 & $+$0.913 & 1.11 & 4.971 & 0.14059 \\
55 & 201104010 & SNR0509-68.7 & $-$200.70 & $-$1.8 & 3.497657 & $-$2.343 & 1.81 & 8.106 & 0.70779 \\
56 & 201096010 & G292\_NW & $-$205.03 & $+$14.2 & 3.497606 & $-$2.394 & 2.174 & 9.735 & 0.18776 \\
57 & 000104000 & Abell2319\_BS & $+$218.72 & $-$12.6 & 3.502554 & $+$2.554 & 1.949 & 8.729 & 0.40517 \\
58 & 201096020 & G292\_NW & $-$204.67 & $+$14.5 & 3.497610 & $-$2.389 & 2.174 & 9.735 & 0.18810 \\
59 & 902001030 & C2025\_N1 & $-$80.00 & $+$26.6 & 3.499066 & $-$0.934 & 1.783 & 7.985 & 0.00594 \\
60 & 000104200 & Abell2319\_BS & $+$218.69 & $-$12.6 & 3.502553 & $+$2.553 & 1.949 & 8.729 & 0.47115 \\
\bottomrule
\end{tabular*}
\end{table*}

\clearpage

\smsection{Data and software availability}{sm:data_avail}

\added{The XRISM/Resolve observations analyzed here (ObsIDs listed
in Table~\ref{tab:obs}) are publicly available through the
JAXA/DARTS and HEASARC archives~\citep{XRISM_DARTS_Public}. The
archival cleaned event lists were produced by the XRISM pipeline
at different epochs (processing versions 03.00.013.009 to
03.03.015.012). The spectra and the response files of all 60
observations were generated uniformly with the XRISM software build
12Sep2025\_V003 distributed with HEASoft~6.36, using the same Resolve
calibration files throughout; the spectral analysis used
XSPEC~12.15.1 with AtomDB/APEC~v3.1.3 atomic data. 
The reduced spectra, response files, and analysis products are available from the authors upon reasonable request.}

\bibliographystyle{apsrev4-2}
\bibliography{refs}

@ARTICLE{2020PhRvD.101l3009M,
       author = {{Malyshev}, Denys and {Thorpe-Morgan}, Charles and {Santangelo}, Andrea and {Jochum}, Josef and {Zhang}, Shuang-Nan},
        title = "{e X T P perspectives for the {\ensuremath{\nu}} MSM sterile neutrino dark matter model}",
      journal = {\prd},
         year = 2020,
        month = jun,
       volume = {101},
       number = {12},
          eid = {123009},
        pages = {123009},
          doi = {10.1103/PhysRevD.101.123009},
archivePrefix = {arXiv},
       eprint = {2001.07014},
 primaryClass = {astro-ph.HE},
       adsurl = {https://ui.adsabs.harvard.edu/abs/2020PhRvD.101l3009M}
}

@ARTICLE{2016PhRvD..93f3518N,
       author = {{Neronov}, A. and {Malyshev}, D.},
        title = "{Toward a full test of the {\ensuremath{\nu}} MSM sterile neutrino dark matter model with Athena}",
      journal = {\prd},
         year = 2016,
        month = mar,
       volume = {93},
       number = {6},
          eid = {063518},
        pages = {063518},
          doi = {10.1103/PhysRevD.93.063518},
archivePrefix = {arXiv},
       eprint = {1509.02758},
 primaryClass = {astro-ph.HE},
       adsurl = {https://ui.adsabs.harvard.edu/abs/2016PhRvD..93f3518N}
}

@ARTICLE{2016A&A...592A.112H,
       author = {{Hofmann}, F. and {Sanders}, J.~S. and {Nandra}, K. and {Clerc}, N. and {Gaspari}, M.},
        title = "{7.1 keV sterile neutrino constraints from X-ray observations of 33 clusters of galaxies with Chandra ACIS}",
      journal = {\aap},
         year = 2016,
        month = aug,
       volume = {592},
          eid = {A112},
        pages = {A112},
          doi = {10.1051/0004-6361/201527977},
archivePrefix = {arXiv},
       eprint = {1606.04091},
 primaryClass = {astro-ph.CO},
       adsurl = {https://ui.adsabs.harvard.edu/abs/2016A&A...592A.112H}
}

@ARTICLE{2016PASJ...68S..31S,
       author = {{Sekiya}, Norio and {Yamasaki}, Noriko Y. and {Mitsuda}, Kazuhisa},
        title = "{A search for a keV signature of radiatively decaying dark matter with Suzaku XIS observations of the X-ray diffuse background}",
      journal = {\pasj},
         year = 2016,
        month = jun,
       volume = {68},
          eid = {S31},
        pages = {S31},
          doi = {10.1093/pasj/psv081},
archivePrefix = {arXiv},
       eprint = {1504.02826},
 primaryClass = {astro-ph.HE},
       adsurl = {https://ui.adsabs.harvard.edu/abs/2016PASJ...68S..31S}
}

@ARTICLE{2014PhRvD..90j3506M,
       author = {{Malyshev}, D. and {Neronov}, A. and {Eckert}, D.},
        title = "{Constraints on 3.55 keV line emission from stacked observations of dwarf spheroidal galaxies}",
      journal = {\prd},
         year = 2014,
        month = nov,
       volume = {90},
       number = {10},
          eid = {103506},
        pages = {103506},
          doi = {10.1103/PhysRevD.90.103506},
archivePrefix = {arXiv},
       eprint = {1408.3531},
 primaryClass = {astro-ph.HE},
       adsurl = {https://ui.adsabs.harvard.edu/abs/2014PhRvD..90j3506M}
}

@article{Dodelson:1993je,
    author        = "Dodelson, Scott and Widrow, Lawrence M.",
    title         = "{Sterile-neutrinos as dark matter}",
    journal       = "Phys. Rev. Lett.",
    volume        = "72",
    pages         = "17--20",
    year          = "1994",
    eprint        = "hep-ph/9303287",
    archivePrefix = "arXiv",
    doi           = "10.1103/PhysRevLett.72.17"
}

@article{Barger:1995ty,
    author        = "Barger, Vernon D. and Phillips, R. J. N. and Sarkar, Subir",
    title         = "{Remarks on the KARMEN anomaly}",
    journal       = "Phys. Lett. B",
    volume        = "352",
    pages         = "365--371",
    year          = "1995",
    eprint        = "hep-ph/9503295",
    archivePrefix = "arXiv",
    doi           = "10.1016/0370-2693(95)00486-5",
    note          = "[Erratum: Phys.Lett.B 356, 617 (1995)]"
}

@article{Bulbul:2014sua,
    author        = "Bulbul, Esra and Markevitch, Maxim and Foster, Adam and Smith, Randall K. and Loewenstein, Michael and Randall, Scott W.",
    title         = "{Detection of An Unidentified Emission Line in the Stacked X-ray spectrum of Galaxy Clusters}",
    journal       = "Astrophys. J.",
    volume        = "789",
    pages         = "13",
    year          = "2014",
    eprint        = "1402.2301",
    archivePrefix = "arXiv",
    primaryClass  = "astro-ph.CO",
    doi           = "10.1088/0004-637X/789/1/13"
}

@article{Boyarsky:2014jta,
    author        = "Boyarsky, Alexey and Ruchayskiy, Oleg and Iakubovskyi, Dmytro and Franse, Jeroen",
    title         = "{Unidentified Line in X-Ray Spectra of the Andromeda Galaxy and Perseus Galaxy Cluster}",
    journal       = "Phys. Rev. Lett.",
    volume        = "113",
    pages         = "251301",
    year          = "2014",
    eprint        = "1402.4119",
    archivePrefix = "arXiv",
    primaryClass  = "astro-ph.CO",
    doi           = "10.1103/PhysRevLett.113.251301"
}

@misc{Boyarsky:2014paa,
    author        = "Boyarsky, A. and Franse, J. and Iakubovskyi, D. and Ruchayskiy, O.",
    title         = "{Comment on the paper `Dark matter searches going bananas: the contribution of Potassium (and Chlorine) to the 3.5 keV line' by T. Jeltema and S. Profumo}",
    year          = "2014",
    eprint        = "1408.4388",
    archivePrefix = "arXiv",
    primaryClass  = "astro-ph.CO",
    howpublished  = "arXiv:1408.4388"
}

@article{Jeltema:2014qfa,
    author        = "Jeltema, Tesla E. and Profumo, Stefano",
    title         = "{Discovery of a 3.5 keV line in the Galactic Centre and a critical look at the origin of the line across astronomical targets}",
    journal       = "Mon. Not. Roy. Astron. Soc.",
    volume        = "450",
    number        = "2",
    pages         = "2143--2152",
    year          = "2015",
    eprint        = "1408.1699",
    archivePrefix = "arXiv",
    primaryClass  = "astro-ph.HE",
    doi           = "10.1093/mnras/stv768"
}

@article{Carlson:2014lla,
    author        = "Carlson, Eric and Jeltema, Tesla and Profumo, Stefano",
    title         = "{Where do the 3.5 keV photons come from? A morphological study of the Galactic Center and of Perseus}",
    journal       = "JCAP",
    volume        = "02",
    pages         = "009",
    year          = "2015",
    eprint        = "1411.1758",
    archivePrefix = "arXiv",
    primaryClass  = "astro-ph.HE",
    doi           = "10.1088/1475-7516/2015/02/009"
}

@article{Riemer-Sorensen:2014yda,
    author        = "Riemer-S{\o}rensen, Signe",
    title         = "{Constraints on the presence of a 3.5 keV dark matter emission line from Chandra observations of the Galactic centre}",
    journal       = "Astron. Astrophys.",
    volume        = "590",
    pages         = "A71",
    year          = "2016",
    eprint        = "1405.7943",
    archivePrefix = "arXiv",
    primaryClass  = "astro-ph.CO",
    doi           = "10.1051/0004-6361/201527278"
}

@ARTICLE{2015arXiv150805186I,
       author = {{Iakubovskyi}, Dmytro and {Bulbul}, Esra and {Foster}, Adam R. and {Savchenko}, Denys and {Sadova}, Valentyna},
        title = "{Testing the origin of \raisebox{-0.5ex}\textasciitilde3.55 keV line in individual galaxy clusters observed with XMM-Newton}",
      journal = {arXiv e-prints},
         year = 2015,
        month = aug,
          eid = {arXiv:1508.05186},
        pages = {arXiv:1508.05186},
          doi = {10.48550/arXiv.1508.05186},
archivePrefix = {arXiv},
       eprint = {1508.05186},
 primaryClass = {astro-ph.HE},
       adsurl = {https://ui.adsabs.harvard.edu/abs/2015arXiv150805186I}
}

@article{XQC:2015mwy,
    author        = "Figueroa-Feliciano, Enectali and others",
    collaboration = "XQC",
    title         = "{Searching for keV Sterile Neutrino Dark Matter with X-ray Microcalorimeter Sounding Rockets}",
    journal       = "Astrophys. J.",
    volume        = "814",
    number        = "1",
    pages         = "82",
    year          = "2015",
    eprint        = "1506.05519",
    archivePrefix = "arXiv",
    primaryClass  = "astro-ph.CO",
    doi           = "10.1088/0004-637X/814/1/82"
}

@article{Iakubovskyi:2014yxa,
    author        = "Iakubovskyi, Dmytro",
    title         = "{New emission line at $\sim$3.5 keV -- observational status, connection with radiatively decaying dark matter and directions for future studies}",
    journal       = "Advances in Astronomy and Space Physics",
    volume        = "4",
    pages         = "9",
    year          = "2014",
    eprint        = "1410.2852",
    archivePrefix = "arXiv",
    primaryClass  = "astro-ph.HE"
}

@article{Abazajian:2017tcc,
    author        = "Abazajian, Kevork N.",
    title         = "{Sterile neutrinos in cosmology}",
    journal       = "Phys. Rept.",
    volume        = "711-712",
    pages         = "1--28",
    year          = "2017",
    eprint        = "1705.01837",
    archivePrefix = "arXiv",
    primaryClass  = "hep-ph",
    doi           = "10.1016/j.physrep.2017.10.003"
}

@article{Boyarsky:2018tvu,
    author        = "Boyarsky, A. and Drewes, M. and Lasserre, T. and Mertens, S. and Ruchayskiy, O.",
    title         = "{Sterile neutrino Dark Matter}",
    journal       = "Prog. Part. Nucl. Phys.",
    volume        = "104",
    pages         = "1--45",
    year          = "2019",
    eprint        = "1807.07938",
    archivePrefix = "arXiv",
    primaryClass  = "hep-ph",
    doi           = "10.1016/j.ppnp.2018.07.004"
}

@article{Hitomi:2016mun,
    author        = "Aharonian, F. A. and others",
    collaboration = "Hitomi",
    title         = "{Hitomi constraints on the 3.5 keV line in the Perseus galaxy cluster}",
    journal       = "Astrophys. J. Lett.",
    volume        = "837",
    number        = "1",
    pages         = "L15",
    year          = "2017",
    eprint        = "1607.07420",
    archivePrefix = "arXiv",
    primaryClass  = "astro-ph.HE",
    doi           = "10.3847/2041-8213/aa61fa"
}

@article{Tashiro:2022ful,
    author  = "Tashiro, Makoto S.",
    title   = "{XRISM: X-ray imaging and spectroscopy mission}",
    journal = "Int. J. Mod. Phys. D",
    volume  = "31",
    number  = "02",
    pages   = "2230001",
    year    = "2022",
    doi     = "10.1142/S0218271822300014"
}

@article{XRISM:2025lzv,
    author        = "Audard, Marc and others",
    collaboration = "XRISM",
    title         = "{XRISM Constraints on Unidentified X-Ray Emission Lines, Including the 3.5 keV Line, in the Stacked Spectrum of 10 Galaxy Clusters}",
    journal       = "Astrophys. J. Lett.",
    volume        = "994",
    number        = "1",
    pages         = "L28",
    year          = "2025",
    eprint        = "2510.24560",
    archivePrefix = "arXiv",
    primaryClass  = "astro-ph.HE",
    doi           = "10.3847/2041-8213/ae17ad"
}

@article{Boyarsky:2006fg,
    author        = "Boyarsky, Alexey and Neronov, A. and Ruchayskiy, O. and Shaposhnikov, M. and Tkachev, I.",
    title         = "{Where to find a dark matter sterile neutrino?}",
    journal       = "Phys. Rev. Lett.",
    volume        = "97",
    pages         = "261302",
    year          = "2006",
    eprint        = "astro-ph/0603660",
    archivePrefix = "arXiv",
    doi           = "10.1103/PhysRevLett.97.261302"
}

@article{Abazajian:2001vt,
    author        = "Abazajian, Kevork and Fuller, George M. and Tucker, Wallace H.",
    title         = "{Direct detection of warm dark matter in the X-ray}",
    journal       = "Astrophys. J.",
    volume        = "562",
    pages         = "593--604",
    year          = "2001",
    eprint        = "astro-ph/0106002",
    archivePrefix = "arXiv",
    doi           = "10.1086/323867"
}

@article{Dessert:2018qih,
    author        = "Dessert, Christopher and Rodd, Nicholas L. and Safdi, Benjamin R.",
    title         = "{The dark matter interpretation of the 3.5-keV line is inconsistent with blank-sky observations}",
    journal       = "Science",
    volume        = "367",
    number        = "6485",
    pages         = "1465--1467",
    year          = "2020",
    eprint        = "1812.06976",
    archivePrefix = "arXiv",
    primaryClass  = "astro-ph.CO",
    doi           = "10.1126/science.aaw3772"
}

@ARTICLE{Boyarsky:2020hqb,
       author = {{Boyarsky}, Alexey and {Malyshev}, Denys and {Ruchayskiy}, Oleg and {Savchenko}, Denys},
        title = "{Technical comment on the paper of Dessert et al. ``The dark matter interpretation of the 3.5 keV line is inconsistent with blank-sky observations''}",
      journal = {SciPost Astronomy Core},
         year = 2021,
        month = may,
       volume = {1},
       number = {1},
          eid = {001},
        pages = {001},
          doi = {10.21468/SciPostAstroCore.1.1.001},
archivePrefix = {arXiv},
       eprint = {2004.06601},
 primaryClass = {astro-ph.CO},
       adsurl = {https://ui.adsabs.harvard.edu/abs/2021ScPAC...1....1B}
}

@misc{Abazajian:2020unr,
    author        = {Abazajian, Kevork N.},
    title         = {Technical Comment on
                     ``The dark matter interpretation of the 3.5-keV line
                     is inconsistent with blank-sky observations''},
    year          = {2020},
    eprint        = {2004.06170},
    archivePrefix = {arXiv},
    primaryClass  = {astro-ph.HE}
}

@article{Dessert:2023fen,
    author        = "Dessert, Christopher and Foster, Joshua W. and Park, Yujin and Safdi, Benjamin R.",
    title         = "{Was There a 3.5 keV Line?}",
    journal       = "Astrophys. J.",
    volume        = "964",
    number        = "2",
    pages         = "185",
    year          = "2024",
    eprint        = "2309.03254",
    archivePrefix = "arXiv",
    primaryClass  = "astro-ph.CO",
    doi           = "10.3847/1538-4357/ad2612"
}

@article{Dessert:2023vyl,
    author        = "Dessert, Christopher and Ning, Orion and Rodd, Nicholas L. and Safdi, Benjamin R.",
    title         = "{Resurrecting Hitomi for Decaying Dark Matter and Forecasting Leading Sensitivity for XRISM}",
    journal       = "Phys. Rev. Lett.",
    volume        = "132",
    number        = "21",
    pages         = "211002",
    year          = "2024",
    eprint        = "2305.17160",
    archivePrefix = "arXiv",
    primaryClass  = "astro-ph.CO",
    doi           = "10.1103/PhysRevLett.132.211002"
}

@article{Gu:2015gqm,
    author        = "Gu, L. and Kaastra, J. and Raassen, A. J. J. and Mullen, P. D. and Cumbee, R. S. and Lyons, D. and Stancil, P. C.",
    title         = "{A novel scenario for the possible X-ray line feature at $\sim$3.5 keV: Charge exchange with bare sulfur ions}",
    journal       = "Astron. Astrophys.",
    volume        = "584",
    pages         = "L11",
    year          = "2015",
    eprint        = "1511.06557",
    archivePrefix = "arXiv",
    primaryClass  = "astro-ph.HE",
    doi           = "10.1051/0004-6361/201527634"
}

@article{Neronov:2016wdd,
    author        = "Neronov, A. and Malyshev, Denys and Eckert, Dominique",
    title         = "{Decaying dark matter search with NuSTAR deep sky observations}",
    journal       = "Phys. Rev. D",
    volume        = "94",
    number        = "12",
    pages         = "123504",
    year          = "2016",
    eprint        = "1607.07328",
    archivePrefix = "arXiv",
    primaryClass  = "astro-ph.HE",
    doi           = "10.1103/PhysRevD.94.123504"
}

@article{Cappelluti:2017ywp,
    author        = "Cappelluti, Nico and Bulbul, Esra and Foster, Adam and Natarajan, Priyamvada and Urry, Megan C. and Bautz, Mark W. and Civano, Francesca and Miller, Eric and Smith, Randall K.",
    title         = "{Searching for the 3.5 keV Line in the Deep Fields with Chandra: the 10 Ms observations}",
    journal       = "Astrophys. J.",
    volume        = "854",
    number        = "2",
    pages         = "179",
    year          = "2018",
    eprint        = "1701.07932",
    archivePrefix = "arXiv",
    primaryClass  = "astro-ph.CO",
    doi           = "10.3847/1538-4357/aaaa68"
}

@ARTICLE{2009arXiv0906.1788D,
       author = {{den Herder}, J.~W. and {Boyarsky}, A. and {Ruchayskiy}, O. and {Abazajian}, K. and {Frenk}, C. and {Hansen}, S. and {Jonker}, P. and {Kouveliotou}, C. and {Lesgourgues}, J. and {Neronov}, A. and {Ohashi}, T. and {Paerels}, F. and {Paltani}, S. and {Piro}, L. and {Pohl}, M. and {Shaposhnikov}, M. and {Silk}, J. and {Valle}, J.},
        title = "{The search for decaying Dark Matter}",
      journal = {arXiv e-prints},
         year = 2009,
        month = jun,
          eid = {arXiv:0906.1788},
        pages = {arXiv:0906.1788},
          doi = {10.48550/arXiv.0906.1788},
archivePrefix = {arXiv},
       eprint = {0906.1788},
 primaryClass = {astro-ph.CO},
       adsurl = {https://ui.adsabs.harvard.edu/abs/2009arXiv0906.1788D}
}

@ARTICLE{2015PhRvL.115p1301B,
       author = {{Boyarsky}, A. and {Franse}, J. and {Iakubovskyi}, D. and {Ruchayskiy}, O.},
        title = "{Checking the Dark Matter Origin of a 3.53 keV Line with the Milky Way Center}",
      journal = {\prl},
         year = 2015,
        month = oct,
       volume = {115},
       number = {16},
          eid = {161301},
        pages = {161301},
          doi = {10.1103/PhysRevLett.115.161301},
archivePrefix = {arXiv},
       eprint = {1408.2503},
 primaryClass = {astro-ph.CO},
       adsurl = {https://ui.adsabs.harvard.edu/abs/2015PhRvL.115p1301B}
}

@ARTICLE{2016ApJ...829..124F,
       author = {{Franse}, Jeroen and {Bulbul}, Esra and {Foster}, Adam and {Boyarsky}, Alexey and {Markevitch}, Maxim and {Bautz}, Mark and {Iakubovskyi}, Dmytro and {Loewenstein}, Mike and {McDonald}, Michael and {Miller}, Eric and {Randall}, Scott W. and {Ruchayskiy}, Oleg and {Smith}, Randall K.},
        title = "{Radial Profile of the 3.5 keV Line Out to R200 in the Perseus Cluster}",
      journal = {\apj},
         year = 2016,
        month = oct,
       volume = {829},
       number = {2},
          eid = {124},
        pages = {124},
          doi = {10.3847/0004-637X/829/2/124},
archivePrefix = {arXiv},
       eprint = {1604.01759},
 primaryClass = {astro-ph.CO},
       adsurl = {https://ui.adsabs.harvard.edu/abs/2016ApJ...829..124F}
}

@ARTICLE{2015MNRAS.451.2447U,
       author = {{Urban}, O. and {Werner}, N. and {Allen}, S.~W. and {Simionescu}, A. and {Kaastra}, J.~S. and {Strigari}, L.~E.},
        title = "{A Suzaku search for dark matter emission lines in the X-ray brightest galaxy clusters}",
      journal = {\mnras},
         year = 2015,
        month = aug,
       volume = {451},
       number = {3},
        pages = {2447-2461},
          doi = {10.1093/mnras/stv1142},
archivePrefix = {arXiv},
       eprint = {1411.0050},
 primaryClass = {astro-ph.CO},
       adsurl = {https://ui.adsabs.harvard.edu/abs/2015MNRAS.451.2447U}
}

@ARTICLE{2018arXiv181210488B,
       author = {{Boyarsky}, A. and {Iakubovskyi}, D. and {Ruchayskiy}, O. and {Savchenko}, D.},
        title = "{Surface brightness profile of the 3.5 keV line in the Milky Way halo}",
      journal = {arXiv e-prints},
         year = 2018,
        month = dec,
          eid = {arXiv:1812.10488},
        pages = {arXiv:1812.10488},
          doi = {10.48550/arXiv.1812.10488},
archivePrefix = {arXiv},
       eprint = {1812.10488},
 primaryClass = {astro-ph.HE},
       adsurl = {https://ui.adsabs.harvard.edu/abs/2018arXiv181210488B}
}

@ARTICLE{2019A&A...625L...7H,
       author = {{Hofmann}, F. and {Wegg}, C.},
        title = "{7.1 keV sterile neutrino dark matter constraints from a deep Chandra X-ray observation of the Galactic bulge Limiting Window}",
      journal = {\aap},
         year = 2019,
        month = may,
       volume = {625},
          eid = {L7},
        pages = {L7},
          doi = {10.1051/0004-6361/201935561},
archivePrefix = {arXiv},
       eprint = {1905.00916},
 primaryClass = {astro-ph.HE},
       adsurl = {https://ui.adsabs.harvard.edu/abs/2019A&A...625L...7H}
}

@article{Foster:2012hy,
    author = "Foster, A. R. and Ji, L. and Smith, R. K. and Brickhouse, N. S.",
    title = "{Updated Atomic Data and Calculations for X-ray Spectroscopy}",
    eprint = "1207.0576",
    archivePrefix = "arXiv",
    primaryClass = "astro-ph.HE",
    doi = "10.1088/0004-637X/756/2/128",
    journal = "Astrophys. J.",
    volume = "756",
    pages = "128",
    year = "2012"
}

@ARTICLE{2002ApJ...576..188M,
       author = {{McCammon}, D. and {Almy}, R. and {Apodaca}, E. and {Bergmann Tiest}, W. and {Cui}, W. and {Deiker}, S. and {Galeazzi}, M. and {Juda}, M. and {Lesser}, A. and {Mihara}, T. and {Morgenthaler}, J.~P. and {Sanders}, W.~T. and {Zhang}, J. and {Figueroa-Feliciano}, E. and {Kelley}, R.~L. and {Moseley}, S.~H. and {Mushotzky}, R.~F. and {Porter}, F.~S. and {Stahle}, C.~K. and {Szymkowiak}, A.~E.},
        title = "{A High Spectral Resolution Observation of the Soft X-Ray Diffuse Background with Thermal Detectors}",
      journal = {\apj},
         year = 2002,
        month = sep,
       volume = {576},
       number = {1},
        pages = {188-203},
          doi = {10.1086/341727},
archivePrefix = {arXiv},
       eprint = {astro-ph/0205012},
 primaryClass = {astro-ph},
       adsurl = {https://ui.adsabs.harvard.edu/abs/2002ApJ...576..188M}
}

@ARTICLE{1982PhRvD..25..766P,
       author = {{Pal}, Palash B. and {Wolfenstein}, Lincoln},
        title = "{Radiative decays of massive neutrinos}",
      journal = {\prd},
         year = 1982,
        month = feb,
       volume = {25},
       number = {3},
        pages = {766-773},
          doi = {10.1103/PhysRevD.25.766},
       adsurl = {https://ui.adsabs.harvard.edu/abs/1982PhRvD..25..766P}
}

@ARTICLE{2001A&A...365L...1J,
       author = {{Jansen}, F. and {Lumb}, D. and {Altieri}, B. and {Clavel}, J. and {Ehle}, M. and {Erd}, C. and {Gabriel}, C. and {Guainazzi}, M. and {Gondoin}, P. and {Much}, R. and {Munoz}, R. and {Santos}, M. and {Schartel}, N. and {Texier}, D. and {Vacanti}, G.},
        title = "{XMM-Newton observatory. I. The spacecraft and operations}",
      journal = {\aap},
         year = 2001,
        month = jan,
       volume = {365},
        pages = {L1-L6},
          doi = {10.1051/0004-6361:20000036},
       adsurl = {https://ui.adsabs.harvard.edu/abs/2001A&A...365L...1J}
}

@ARTICLE{2007PASJ...59S...1M,
       author = {{Mitsuda}, Kazuhisa and others},
        title = "{The X-Ray Observatory Suzaku}",
      journal = {\pasj},
         year = 2007,
        month = jan,
       volume = {59},
        pages = {S1-S7},
          doi = {10.1093/pasj/59.sp1.S1},
       adsurl = {https://ui.adsabs.harvard.edu/abs/2007PASJ...59S...1M}
}

@INPROCEEDINGS{2000SPIE.4012....2W,
       author = {{Weisskopf}, Martin C. and {Tananbaum}, Harvey D. and {Van Speybroeck}, Leon P. and {O'Dell}, Stephen L.},
        title = "{Chandra X-ray Observatory (CXO): overview}",
    booktitle = {X-Ray Optics, Instruments, and Missions III},
         year = 2000,
       editor = {{Truemper}, Joachim E. and {Aschenbach}, Bernd},
       series = {Society of Photo-Optical Instrumentation Engineers (SPIE) Conference Series},
       volume = {4012},
        month = jul,
        pages = {2-16},
          doi = {10.1117/12.391545},
archivePrefix = {arXiv},
       eprint = {astro-ph/0004127},
 primaryClass = {astro-ph},
       adsurl = {https://ui.adsabs.harvard.edu/abs/2000SPIE.4012....2W}
}

@ARTICLE{2020arXiv200304962X,
       author = {{XRISM Science Team}},
        title = "{Science with the X-ray Imaging and Spectroscopy Mission (XRISM)}",
      journal = {arXiv e-prints},
         year = 2020,
        month = mar,
          eid = {arXiv:2003.04962},
        pages = {arXiv:2003.04962},
          doi = {10.48550/arXiv.2003.04962},
archivePrefix = {arXiv},
       eprint = {2003.04962},
 primaryClass = {astro-ph.HE},
       adsurl = {https://ui.adsabs.harvard.edu/abs/2020arXiv200304962X}
}

@misc{xrism_pog,
  title = {{XRISM} Proposers' Observatory Guide},
  author = {{XRISM Team}},
  year = {2024},
  howpublished = {\url{https://heasarc.gsfc.nasa.gov/docs/xrism/proposals/POG/}},
  note = {Accessed: 2026}
}

@ARTICLE{2016ARA&A..54..529B,
       author = {{Bland-Hawthorn}, Joss and {Gerhard}, Ortwin},
        title = "{The Galaxy in Context: Structural, Kinematic, and Integrated Properties}",
      journal = {\araa},
         year = 2016,
        month = sep,
       volume = {54},
        pages = {529-596},
          doi = {10.1146/annurev-astro-081915-023441},
archivePrefix = {arXiv},
       eprint = {1602.07702},
 primaryClass = {astro-ph.GA},
       adsurl = {https://ui.adsabs.harvard.edu/abs/2016ARA&A..54..529B}
}

@ARTICLE{1996ApJ...462..563N,
       author = {{Navarro}, Julio F. and {Frenk}, Carlos S. and {White}, Simon D.~M.},
        title = "{The Structure of Cold Dark Matter Halos}",
      journal = {\apj},
         year = 1996,
        month = may,
       volume = {462},
        pages = {563},
          doi = {10.1086/177173},
archivePrefix = {arXiv},
       eprint = {astro-ph/9508025},
 primaryClass = {astro-ph},
       adsurl = {https://ui.adsabs.harvard.edu/abs/1996ApJ...462..563N}
}

@ARTICLE{2010JCAP...08..004C,
       author = {{Catena}, Riccardo and {Ullio}, Piero},
        title = "{A novel determination of the local dark matter density}",
      journal = {\jcap},
         year = 2010,
        month = aug,
       volume = {2010},
       number = {8},
          eid = {004},
        pages = {004},
          doi = {10.1088/1475-7516/2010/08/004},
archivePrefix = {arXiv},
       eprint = {0907.0018},
 primaryClass = {astro-ph.CO},
       adsurl = {https://ui.adsabs.harvard.edu/abs/2010JCAP...08..004C}
}

@ARTICLE{2013JCAP...07..016N,
       author = {{Nesti}, Fabrizio and {Salucci}, Paolo},
        title = "{The Dark Matter halo of the Milky Way, AD 2013}",
      journal = {\jcap},
         year = 2013,
        month = jul,
       volume = {2013},
       number = {7},
          eid = {016},
        pages = {016},
          doi = {10.1088/1475-7516/2013/07/016},
archivePrefix = {arXiv},
       eprint = {1304.5127},
 primaryClass = {astro-ph.GA},
       adsurl = {https://ui.adsabs.harvard.edu/abs/2013JCAP...07..016N}
}

@ARTICLE{2010MNRAS.403.1829S,
       author = {{Sch{\"o}nrich}, Ralph and {Binney}, James and {Dehnen}, Walter},
        title = "{Local kinematics and the local standard of rest}",
      journal = {\mnras},
         year = 2010,
        month = apr,
       volume = {403},
       number = {4},
        pages = {1829-1833},
          doi = {10.1111/j.1365-2966.2010.16253.x},
archivePrefix = {arXiv},
       eprint = {0912.3693},
 primaryClass = {astro-ph.GA},
       adsurl = {https://ui.adsabs.harvard.edu/abs/2010MNRAS.403.1829S}
}

@ARTICLE{2011MNRAS.414.2446M,
       author = {{McMillan}, Paul J.},
        title = "{Mass models of the Milky Way}",
      journal = {\mnras},
         year = 2011,
        month = jul,
       volume = {414},
       number = {3},
        pages = {2446-2457},
          doi = {10.1111/j.1365-2966.2011.18564.x},
archivePrefix = {arXiv},
       eprint = {1102.4340},
 primaryClass = {astro-ph.GA},
       adsurl = {https://ui.adsabs.harvard.edu/abs/2011MNRAS.414.2446M}
}

@ARTICLE{2021PDU....3200826B,
       author = {{Benito}, Mar{\'\i}a and {Iocco}, Fabio and {Cuoco}, Alessandro},
        title = "{Uncertainties in the Galactic Dark Matter distribution: An update}",
      journal = {Physics of the Dark Universe},
         year = 2021,
        month = may,
       volume = {32},
          eid = {100826},
        pages = {100826},
          doi = {10.1016/j.dark.2021.100826},
archivePrefix = {arXiv},
       eprint = {2009.13523},
 primaryClass = {astro-ph.GA},
       adsurl = {https://ui.adsabs.harvard.edu/abs/2021PDU....3200826B}
}

@ARTICLE{1967RvMP...39...78B,
       author = {{Bearden}, J.~A.},
        title = "{X-Ray Wavelengths}",
      journal = {Reviews of Modern Physics},
         year = 1967,
        month = jan,
       volume = {39},
       number = {1},
        pages = {78-124},
          doi = {10.1103/RevModPhys.39.78},
       adsurl = {https://ui.adsabs.harvard.edu/abs/1967RvMP...39...78B}
}

@misc{KortrightThompson:XDB,
  author       = {Kortright, Jeffrey B. and Thompson, Albert C.},
  title        = {{X-Ray Data Booklet, Section 1.2: X-Ray Emission Energies, Table 1-3}},
  howpublished = {Lawrence Berkeley National Laboratory, Center for X-ray Optics and Advanced Light Source},
  note         = {Photon energies and relative intensities of K-, L-, and M-shell lines},
  url          = {https://xdb.lbl.gov/Section1/Table_1-3.pdf}
}

@ARTICLE{2017MNRAS.465...76M,
       author = {{McMillan}, Paul J.},
        title = "{The mass distribution and gravitational potential of the Milky Way}",
      journal = {\mnras},
         year = 2017,
        month = feb,
       volume = {465},
       number = {1},
        pages = {76-94},
          doi = {10.1093/mnras/stw2759},
archivePrefix = {arXiv},
       eprint = {1608.00971},
 primaryClass = {astro-ph.GA},
       adsurl = {https://ui.adsabs.harvard.edu/abs/2017MNRAS.465...76M}
}

@ARTICLE{2020MNRAS.494.4291C,
       author = {{Cautun}, Marius and {Ben{\'\i}tez-Llambay}, Alejandro and {Deason}, Alis J. and {Frenk}, Carlos S. and {Fattahi}, Azadeh and {G{\'o}mez}, Facundo A. and {Grand}, Robert J.~J. and {Oman}, Kyle A. and {Navarro}, Julio F. and {Simpson}, Christine M.},
        title = "{The milky way total mass profile as inferred from Gaia DR2}",
      journal = {\mnras},
         year = 2020,
        month = may,
       volume = {494},
       number = {3},
        pages = {4291-4313},
          doi = {10.1093/mnras/staa1017},
archivePrefix = {arXiv},
       eprint = {1911.04557},
 primaryClass = {astro-ph.GA},
       adsurl = {https://ui.adsabs.harvard.edu/abs/2020MNRAS.494.4291C}
}

@ARTICLE{2021RPPh...84j4901D,
       author = {{de Salas}, Pablo F. and {Widmark}, Axel},
        title = "{Dark matter local density determination: recent observations and future prospects}",
      journal = {Reports on Progress in Physics},
         year = 2021,
        month = oct,
       volume = {84},
       number = {10},
          eid = {104901},
        pages = {104901},
          doi = {10.1088/1361-6633/ac24e7},
archivePrefix = {arXiv},
       eprint = {2012.11477},
 primaryClass = {astro-ph.GA},
       adsurl = {https://ui.adsabs.harvard.edu/abs/2021RPPh...84j4901D}
}

@ARTICLE{PDG2025,
       author = {{Navas}, S. and others},
 collaboration = {Particle Data Group},
        title = "{Review of Particle Physics}",
      journal = {Phys. Rev. D},
         year = 2024,
        month = aug,
       volume = {110},
       number = {3},
          eid = {030001},
        pages = {030001},
          doi = {10.1103/PhysRevD.110.030001},
       adsurl = {https://ui.adsabs.harvard.edu/abs/2024PhRvD.110c0001N}
}

@misc{XRISMMASTR,
  title        = {{XRISMMASTR}: XRISM Master Catalog},
  author       = {{HEASARC}},
  howpublished = {\url{https://heasarc.gsfc.nasa.gov/W3Browse/catalog/xrismmastr.html}},
  note         = {Accessed 2026},
  year         = {2026}
}

@misc{XRISM_DARTS_Public,
  title        = {{XRISM} Public Data List},
  author       = {{DARTS, ISAS/JAXA}},
  howpublished = {\url{https://data.darts.isas.jaxa.jp/pub/xrism/browse/public_list/}},
  note         = {Accessed 2026},
  year         = {2026}
}

@misc{XRISM_LongTermPlan,
  title        = {{XRISM} Long-term Observation Plan},
  author       = {{XRISM Science Operations Team}},
  howpublished = {\url{https://xrism.isas.jaxa.jp/research/observers/longterm/index.html}},
  note         = {Accessed 2026},
  year         = {2026}
}

@misc{XRISM_AO3_Targets,
  title        = {{XRISM} GO Cycle 3 Approved Target List},
  author       = {{HEASARC}},
  howpublished = {\url{https://heasarc.gsfc.nasa.gov/docs/xrism/proposals/ao3/AO3_approved_targets.html}},
  note         = {Accessed 2026},
  year         = {2026}
}

@ARTICLE{2023ApJ...946...73Z,
       author = {{Zhou}, Yuan and {Li}, Xinyi and {Huang}, Yang and {Zhang}, Huawei},
        title = "{The Circular Velocity Curve of the Milky Way from 5-25 kpc Using Luminous Red Giant Branch Stars}",
      journal = {\apj},
         year = 2023,
        month = apr,
       volume = {946},
       number = {2},
          eid = {73},
        pages = {73},
          doi = {10.3847/1538-4357/acadd9},
archivePrefix = {arXiv},
       eprint = {2212.10393},
 primaryClass = {astro-ph.GA},
       adsurl = {https://ui.adsabs.harvard.edu/abs/2023ApJ...946...73Z}
}

@ARTICLE{2015PASJ...67...23T,
       author = {{Tamura}, Takayuki and {Iizuka}, Ryo and {Maeda}, Yoshitomo and {Mitsuda}, Kazuhisa and {Yamasaki}, Noriko Y.},
        title = "{An X-ray spectroscopic search for dark matter in the Perseus cluster with Suzaku}",
      journal = {\pasj},
         year = 2015,
        month = apr,
       volume = {67},
       number = {2},
          eid = {23},
        pages = {23},
          doi = {10.1093/pasj/psu156},
archivePrefix = {arXiv},
       eprint = {1412.1869},
 primaryClass = {astro-ph.HE},
       adsurl = {https://ui.adsabs.harvard.edu/abs/2015PASJ...67...23T}
}

@ARTICLE{2014PhRvD..89b5017H,
       author = {{Horiuchi}, Shunsaku and {Humphrey}, Philip J. and {O{\~n}orbe}, Jose and {Abazajian}, Kevork N. and {Kaplinghat}, Manoj and {Garrison-Kimmel}, Shea},
        title = "{Sterile neutrino dark matter bounds from galaxies of the Local Group}",
      journal = {\prd},
         year = 2014,
        month = jan,
       volume = {89},
       number = {2},
          eid = {025017},
        pages = {025017},
          doi = {10.1103/PhysRevD.89.025017},
archivePrefix = {arXiv},
       eprint = {1311.0282},
 primaryClass = {astro-ph.CO},
       adsurl = {https://ui.adsabs.harvard.edu/abs/2014PhRvD..89b5017H}
}

@misc{XRISM_ABC_Guide,
  title        = {The {XRISM} Data Reduction Guide ({ABC} Guide), Version 2.0},
  author       = {{XRISM Science Data Center} and {XRISM Guest Observer Facility}},
  howpublished = {\url{https://heasarc.gsfc.nasa.gov/docs/xrism/analysis/abc_guide/}},
  note         = {Accessed 2026}
}

@inproceedings{Doyle2022_pipeline,
  author       = {Doyle, T. F. and Holland, M. P. and Hill, R. S. and Yaqoob, T. and Loewenstein, M. and Miller, E. D. and others},
  title        = {{The XRISM pipeline software system: connecting continents, processes, testing, and scientists}},
  booktitle    = {Software and Cyberinfrastructure for Astronomy VII},
  series       = {Proc. SPIE},
  volume       = {12181},
  pages        = {1218160},
  year         = {2022},
  doi          = {10.1117/12.2629103},
  eprint       = {2210.07840},
  archivePrefix= {arXiv}
}

@article{XRISM:2025pca,
    author = "Audard, Marc and others",
    collaboration = "XRISM",
    title = "{Structured ionized winds shooting out from a quasar at relativistic speeds}",
    eprint = "2505.09171",
    archivePrefix = "arXiv",
    primaryClass = "astro-ph.HE",
    doi = "10.1038/s41586-025-08968-2",
    journal = "Nature",
    volume = "641",
    number = "8065",
    pages = "1132--1136",
    year = "2025"
}

@ARTICLE{2016ApJ...833...31B,
       author = {{Bovy}, Jo and {Bahmanyar}, Anita and {Fritz}, Tobias K. and {Kallivayalil}, Nitya},
        title = "{\textcolor{blue}{The Shape of the Inner Milky Way Halo from Observations of the Pal 5 and GD-1 Stellar Streams}}",
      journal = {\apj},
         year = 2016,
        month = dec,
       volume = {833},
       number = {1},
        pages = {31},
       eprint = {1609.01298},
archivePrefix = {arXiv},
 primaryClass = {astro-ph.GA},
       adsurl = {https://ui.adsabs.harvard.edu/abs/2016ApJ...833...31B}
}

@ARTICLE{2019MNRAS.485.3296W,
       author = {{Wegg}, Christopher and {Gerhard}, Ortwin and {Bieth}, Marie},
        title = "{\textcolor{blue}{The gravitational force field of the Galaxy measured from the kinematics of RR Lyrae in Gaia}}",
      journal = {\mnras},
         year = 2019,
        month = may,
       volume = {485},
       number = {3},
        pages = {3296-3316},
       eprint = {1806.09635},
archivePrefix = {arXiv},
 primaryClass = {astro-ph.GA},
       adsurl = {https://ui.adsabs.harvard.edu/abs/2019MNRAS.485.3296W}
}

@INPROCEEDINGS{Xspec,
       author = {{Arnaud}, K.~A.},
        title = "{XSPEC: The First Ten Years}",
    booktitle = {Astronomical Data Analysis Software and Systems V},
         year = 1996,
       editor = {{Jacoby}, George H. and {Barnes}, Jeannette},
       series = {Astronomical Society of the Pacific Conference Series},
       volume = {101},
        pages = {17},
       adsurl = {https://ui.adsabs.harvard.edu/abs/1996ASPC..101...17A}
}

@INPROCEEDINGS{2025AAS...24611804F,
       author = {{Foster}, Adam R. and others},
        title = "{AtomDB 3.1: Updated Spectral Data for X-ray Analysis}",
    booktitle = {American Astronomical Society Meeting Abstracts},
         year = 2025,
       series = {American Astronomical Society Meeting Abstracts},
       volume = {246},
          eid = {118.04},
        pages = {118.04},
       adsurl = {https://ui.adsabs.harvard.edu/abs/2025AAS...24611804F}
}

@ARTICLE{2025JATIS..11d2016P,
       author = {{Porter}, Frederick S. and {Kilbourne}, Caroline A. and {Chiao}, Meng and {Cumbee}, Renata and {Eckart}, Megan E. and {Fujimoto}, Ryuichi and {Ishisaki}, Yoshitaka and {Kanemaru}, Yoshiaki and {Kelley}, Richard L. and {Leutenegger}, Maurice and {Maeda}, Yoshitomo and {Mizumoto}, Misaki and {Sato}, Kosuke and {Sawada}, Makoto and {Sneiderman}, Gary and {Takei}, Yoh and {Tsujimoto}, Masahiro and {Uchida}, Yuusuke and {Watanabe}, Tomomi and {Yamada}, Shinya},
        title = "{In-flight performance of the XRISM/Resolve detector system}",
      journal = {Journal of Astronomical Telescopes, Instruments, and Systems},
         year = 2025,
        month = oct,
       volume = {11},
          eid = {042016},
        pages = {042016},
          doi = {10.1117/1.JATIS.11.4.042016},
       adsurl = {https://ui.adsabs.harvard.edu/abs/2025JATIS..11d2016P}
}

@article{Kelley:2025resolve,
    author = "Kelley, Richard L. and Ishisaki, Yoshitaka and Costantini, Elisa and others",
    title = "{Resolve instrument onboard the X-Ray Imaging and Spectroscopy Mission}",
    journal = "J. Astron. Telesc. Instrum. Syst.",
    volume = "11",
    number = "4",
    pages = "042026",
    year = "2025",
    doi = "10.1117/1.JATIS.11.4.042026"
}

@article{Sicilian:2020glg,
    author = "Sicilian, Dominic and Cappelluti, Nico and Bulbul, Esra and Civano, Francesca and Moscetti, Massimo and Reynolds, Christopher S.",
    title = "{Probing the Milky Way{\textquoteright}s Dark Matter Halo for the 3.5 keV Line}",
    eprint = "2008.02283",
    archivePrefix = "arXiv",
    primaryClass = "astro-ph.HE",
    doi = "10.3847/1538-4357/abbee9",
    journal = "Astrophys. J.",
    volume = "905",
    number = "2",
    pages = "146",
    year = "2020"
}

@article{Zhou:2024sto,
    author = "Zhou, Yu and Takhistov, Volodymyr and Mitsuda, Kazuhisa",
    title = "{Unlocking Discovery Potential for Decaying Dark Matter and Faint X-Ray Sources with XRISM}",
    eprint = "2407.18189",
    archivePrefix = "arXiv",
    primaryClass = "astro-ph.HE",
    doi = "10.3847/1538-4357/ad83cf",
    journal = "Astrophys. J.",
    volume = "976",
    number = "2",
    pages = "238",
    year = "2024"
}

@ARTICLE{2019ApJ...871..120E,
       author = {{Eilers}, Anna-Christina and {Hogg}, David W. and {Rix}, Hans-Walter and {Ness}, Melissa K.},
        title = "{The Circular Velocity Curve of the Milky Way from 5 to 25 kpc}",
      journal = {The Astrophysical Journal},
         year = 2019,
        month = jan,
       volume = {871},
       number = {1},
        pages = {120},
          doi = {10.3847/1538-4357/aaf648},
       eprint = {1810.09466},
       adsurl = {https://ui.adsabs.harvard.edu/abs/2019ApJ...871..120E}
}

@ARTICLE{1997MNRAS.285..449C,
       author = {{Chen}, L. -W. and {Fabian}, A. C. and {Gendreau}, K. C.},
        title = "{ASCA and ROSAT observations of the QSF3 field: the X-ray background in the 0.1-7 keV band}",
      journal = {Monthly Notices of the Royal Astronomical Society},
         year = 1997,
        month = mar,
       volume = {285},
       number = {3},
        pages = {449-471},
          doi = {10.1093/mnras/285.3.449},
       eprint = {astro-ph/9511089},
       adsurl = {https://ui.adsabs.harvard.edu/abs/1997MNRAS.285..449C}
}

@article{Ota:2026nsr,
    author = "Ota, Naomi and others",
    title = "{Bulk and turbulent gas motions in the interacting galaxy cluster Abell 3395 South observed with XRISM}",
    eprint = "2602.21580",
    archivePrefix = "arXiv",
    primaryClass = "astro-ph.CO",
    doi = "10.1093/pasj/psag079",
    journal = "Publ. Astron. Soc. Jap.",
    volume = "78",
    number = "4",
    pages = "1546--1554",
    year = "2026"
}

@article{Simionescu:2011ii,
    author = "Simionescu, Aurora and others",
    title = "{Baryons at the Edge of the X-ray Brightest Galaxy Cluster}",
    eprint = "1102.2429",
    archivePrefix = "arXiv",
    primaryClass = "astro-ph.CO",
    doi = "10.1126/science.1200331",
    journal = "Science",
    volume = "331",
    pages = "1576",
    year = "2011"
}

@article{Piffaretti:2010my,
    author = "Piffaretti, R. and Arnaud, M. and Pratt, G. W. and Pointecouteau, E. and Melin, J. -B.",
    title = "{The MCXC: a Meta-Catalogue of X-ray detected Clusters of galaxies}",
    eprint = "1007.1916",
    archivePrefix = "arXiv",
    primaryClass = "astro-ph.CO",
    doi = "10.1051/0004-6361/201015377",
    journal = "Astron. Astrophys.",
    volume = "534",
    pages = "A109",
    year = "2011"
}

@misc{SM,
  note = "See Supplemental Material at [URL will be inserted by publisher] for the selection of the observations, the Milky Way halo model and the Doppler corrections, the fitting procedures, the analysis of the stacked spectrum and of the $3.531$~keV residual, and the sky distribution of the excess photons, which includes Refs.~\cite{Protassov:2002sz,1996ApJ...462..563N,2010JCAP...08..004C,2010MNRAS.403.1829S,2011MNRAS.414.2446M,2016ARA&A..54..529B,2016ApJ...833...31B,2019ApJ...871..120E,2019MNRAS.485.3296W,2021PDU....3200826B,2021RPPh...84j4901D,2023ApJ...946...73Z,Ota:2026nsr,Piffaretti:2010my,Simionescu:2011ii}."
}

@article{Protassov:2002sz,
    author = "Protassov, Rostislav and van Dyk, David A. and Connors, Alanna and Kashyap, Vinay L. and Siemiginowska, Aneta",
    title = "{Statistics: handle with care, detecting multiple model components with the likelihood ratio test}",
    eprint = "astro-ph/0201547",
    archivePrefix = "arXiv",
    doi = "10.1086/339856",
    journal = "Astrophys. J.",
    volume = "571",
    pages = "545",
    year = "2002"
}

\end{document}